\documentclass[aip,pop,reprint]{revtex4-1}
\usepackage{amssymb}
\usepackage{graphicx}

\newcommand{\betavsq}{\frac{m_e \beta v^2}{2}}

\newcommand{\Lnhalf}{L_n^{\left(\frac{1}{2}\right)}}
\newcommand{\Lkhalf}{L_k^{\left(\frac{1}{2}\right)}}
\newcommand{\Llhalf}{L_l^{\left(\frac{1}{2}\right)}}
\newcommand{\Lmhalf}{L_m^{\left(\frac{1}{2}\right)}}

\newcommand{\bfk}{{\bf k}}
\newcommand{\bfv}{{\bf v}}
\newcommand{\bfvpr}{{\bf v}'}

\newcommand{\epssqquant}{\left|\epsilon \left(k,\bfk \cdot \bfv + \frac{\hbar k^2}{2 m} \right) \right|^2}

\newcommand{\Lminhalf}{L^{\left(-\frac{1}{2}\right)}}

\newcommand{\nmax}{n_\mathrm{max}}

\begin{document}


\title{Numerical solution of the quantum Lenard-Balescu equation for a one-component plasma}


\author{Christian R. Scullard}
\affiliation{Lawrence Livermore National Laboratory, Livermore CA 94550, USA}
\author{Andrew P. Belt}
\altaffiliation[present address:]{University of Tennessee, Knoxville TN 37996, USA}
\affiliation{Institute for Pure and Applied Mathematics, UCLA, Los Angeles CA 90095, USA}
\author{Susan C. Fennell}
\altaffiliation[present address:]{University of Limerick, Limerick, Ireland}
\affiliation{Institute for Pure and Applied Mathematics, UCLA, Los Angeles CA 90095, USA}
\author{Marija R. Jankovi\'c}
\altaffiliation[present address:]{University of Belgrade, Studentski Trg 12, 11000 Belgrade, Serbia}
\affiliation{Institute for Pure and Applied Mathematics, UCLA, Los Angeles CA 90095, USA}
\author{Nathan Ng}
\altaffiliation[present address:]{University of Maryland, College Park, MD 20742}
\affiliation{Institute for Pure and Applied Mathematics, UCLA, Los Angeles CA 90095, USA}
\author{Susana Serna}
\affiliation{Departament de Matematiques, Universitat Aut\`{o}noma de Barcelona, 08193 Bellaterra-Barcelona, Spain}
\author{Frank R. Graziani}
\affiliation{Lawrence Livermore National Laboratory, Livermore CA 94550, USA}
\email[]{scullard1@llnl.gov}


\date{\today}

\begin{abstract}
We present a numerical solution of the quantum Lenard-Balescu equation using a spectral method, namely an expansion in Laguerre polynomials. This method exactly conserves both particles and energy and facilitates the integration over the dielectric function. To demonstrate the method, we solve the equilibration problem for a spatially homogeneous one-component plasma with various initial conditions. Unlike the more usual Landau/Fokker-Planck system, this method requires no input Coulomb logarithm; the logarithmic terms in the collision integral arise naturally from the equation along with the non-logarithmic order-unity terms. The spectral method can also be used to solve the Landau equation and a quantum version of the Landau equation in which the integration over the wavenumber requires only a lower cutoff. We solve these problems as well and compare them with the full Lenard-Balescu solution in the weak-coupling limit. Finally, we discuss the possible generalization of this method to include spatial inhomogeneity and velocity anisotropy.
\end{abstract}

\pacs{}

\maketitle

\section{Introduction}
The Landau equation, or its equivalent formulation in terms of the Fokker-Planck equation \cite{Rosenbluth}, is a valuable tool in the study of out-of-equilibrium weakly-coupled plasmas \cite{Cohen,Spitzer}. The assumption of small-angle binary scattering between the particles is well-suited to Coulomb interactions at high temperature and low density. However, this approximation results in a divergence at small impact parameters, and the neglect of screening leads to a divergence for large particle separations due to the long-range nature of the Coulomb interaction. As is well-known, these divergences require cutoffs which in practice means choosing a Coulomb logarithm and thereby adding a level of ambiguity to the calculation. Although a more realistic calculation does contain such a logarithmic term, there are other terms potentially the same order as $\log \Lambda$ that we are discarding by using Landau/Fokker-Planck. To include these terms requires a more sophisticated collision operator. One candidate is the quantum Lenard-Balescu (QLB) equation, which accounts for both quantum diffraction and dynamical screening in a natural way and thus requires no input Coulomb logarithm. This equation has been used extensively to calculate various plasma properties at weak coupling, such as transport coefficients \cite{Williams,Whitley,Morales,Ichimaru} and temperature equilibration rates \cite{Daligault,Benedict}. These computations do not require a time-dependent solution of the QLB equation, and indeed the latter has rarely been attempted; the quantum Lenard-Balescu equation is far more complicated than Landau/Fokker-Planck, which itself is not trivial to solve \cite{Chang,Epperlein}. We present here a numerical solution of the quantum Lenard-Balescu equation for a velocity-isotropic, spatially homogeneous, one-component plasma. 

The paper is organized as follows. In the next section we describe the equation in detail and in section \ref{sec:method} we introduce our solution method, which, for reasons discussed there, is very different from those traditionally used to solve the Fokker-Planck equation. In sections \ref{sec:dielectric} and \ref{sec:coeff} we describe how we solve the most difficult problem, the integration over the dielectric function. Our solution method can easily be applied to several simpler kinetic equations, such as the Landau equation, and we enumerate these in section \ref{sec:special} and give the minor modifications needed for each. In sections \ref{sec:IC} to \ref{sec:results} we describe our initial conditions, the numerical solution of the ordinary differential equations that arise from our method, and we show the relaxation to equilibrium of various initial distributions. In the remainder of the paper, we discuss possible generalizations of the method to handle anistropy in velocity and inhomogeneity in space.

\section{Quantum Lenard-Balescu equation}
The equation we will solve is the non-degenerate quantum Lenard-Balescu equation for a one-component plasma,
\begin{equation}
 \frac{\partial f}{\partial t}=C_{QLB}(f) \label{eq:QLB}
\end{equation}
where
\begin{eqnarray}
C_{QLB}(f)=&-&\frac{1}{4 \pi^2 \hbar^2} \int d^3 {\bf v'} \int d^3 {\bf k} \frac{|\phi(k)|^2}{\epssqquant} \cr
\times\delta[\bfk \cdot (\bfv-\bfvpr)&+&\hbar k^2/m][f(\bfv) f(\bfvpr) \cr
&-& f(\bfv+\hbar \bfk/m)f(\bfvpr-\hbar \bfk/m)] \ , \label{eq:Cee}
\end{eqnarray}
where $m$ is the particle's mass, $\hbar$ is Planck's constant, and we use the Coulomb potential,
\begin{equation}
 \phi(k)=\frac{4 \pi e^2}{k^2} .
\end{equation}
The dielectric function is given by,
\begin{equation}
\epsilon(k,\omega)=1-\frac{4 \pi e^2}{k^2}\chi(k,\omega)  \label{eq:Qdelectric} 
\end{equation}
where, in the random phase approximation, the response function, $\chi(k,\omega)$, is given by free-particle expression,
\begin{equation}
 \chi(k,\omega) = \lim_{\eta \rightarrow 0^+} \int d^3 {\bfv} \frac{f(\bfv)-f(\bfv+\hbar \bfk/m)}{\hbar \omega - \hbar \bfv \cdot \bfk - \frac{\hbar^2 k^2}{2 m}+i \eta} . \label{eq:chi}
\end{equation} 
This equation is valid when the system is non-degenerate, i.e., when 
\begin{equation}
 \theta \equiv \frac{2 m k_B T}{\hbar^2 (3 \pi^2 n)^{2/3}} \gg 1 \ ,
\end{equation}
where $n$ is the number density, and weakly-coupled,
\begin{equation}
 \Gamma \equiv \frac{e^2 (4/3 \pi n)^{1/3}}{k_B T} \ll 1 \ .
\end{equation}
When the former condition is violated, additional factors of $1-f$ appear in the integrand in (\ref{eq:Cee}), and the latter is required for the validity of the random phase approximation. Generally speaking, this equation describes high-temperature, low-density plasmas.

The presence of the distribution in the response function is a serious complication. Even worse, integrals over the dielectric function often contain very narrow peaks and their numerical integration can be tricky even at equilibrium \cite{Vorberger,Chapman} let alone for arbitrary distributions. These difficulties, coupled with the fact that the Landau equation, despite its deficiencies, yields distributions that are likely qualitatively correct at weak coupling, have kept the Lenard-Balescu equation from being studied numerically in any serious way. We know of only one previous attempt: Dolinsky's pioneering 1965 solution of the classical LB equation \cite{Dolinsky} using a discretization method in velocity. This work predates the advent of conservative velocity discretization schemes even for the Fokker-Planck equation, but it is not completely clear that such methods are generically well-suited to the Lenard-Balescu equation anway because of the need to accurately integrate over the features of the dielectric function. This issue could certainly use a more thorough investigation. In any case the classical equation considered by Dolinsky is divergent at large $k$ and, unlike the quantum version, an artificial cutoff is needed. Besides Dolinksy, we know only of the somewhat related work of Ricci and Lapenta \cite{Ricci}, in which they consider a one-dimensional version of the Lenard-Balescu equation. While certainly interesting, their system is primarily of theoretical value (it cannot equilibrate, for example). Although many sophisticated techniques are now available for the Landau and Boltzmann equations \cite{Tzoufras,Taitano2015,Taitano2015-2,Gamba2014,Haack2012,Bobylev}, enabling solution in multiple spatial and velocity dimensions with several different particle species, we are only capable, for the moment, of a solution of the QLB equation for a spatially homogeneous, one-component plasma with an isotropic velocity distribution. In section \ref{sec:generalizations}, we will discuss how the method can be generalized. 

As we explain in detail in section \ref{sec:coeff}, after the initial condition has been chosen, only one dimensionless combination of the various physical parameters is really important in the subsequent evolution. We therefore do not lose much by specializing to electrons, so that $m$ in the above equations is equal to the electron mass, $m_e$, and fixing the number density which we shall henceforth call $n_e$. 

\section{Method} \label{sec:method}
Because of the difficulties associated with the dielectric function, we choose to steer clear of discretization in velocity. Instead we use an expansion in Laguerre polynomials,
\begin{equation}
f(v,t)=f^{\mathrm{eq}}(v) \sum_{n=0}^{\infty} A_n(t) \Lnhalf \left(\frac{u \beta m_e v^2}{2} \right) \label{eq:expansion}
\end{equation}
where
\begin{equation}
 f^{\mathrm{eq}}(v) \equiv n_e \left(\frac{m_e \beta}{2 \pi}\right)^{3/2} \exp \left(-\betavsq \right)
\end{equation}
is the Maxwell distribution, $\beta \equiv 1/k_B T$, $k_B$ is Boltzmann's constant, $T$ the temperature of the final equilibrium state, and $n_e$ is the particle number density. The parameter $u \in [1,2]$ will be discussed in detail below. Multiplying by the Maxwell distribution is convenient because it is the stationary solution of this form of the QLB equation, and thus in equilibrium we will simply have
\begin{equation}
 A_n=\delta_{n0} .
\end{equation}
In other words, the action of the collision operator is to attempt to drive down all coefficients with $n>0$. Because we are multiplying by the Maxwell distribution, the Laguerre orthogonality property proves useful
\begin{equation}
  \int_0^\infty x^\alpha e^{-x} L_n^{(\alpha)}(x)L_m^{(\alpha)}(x)dx=\frac{\Gamma(n+\alpha+1)}{n!} \delta_{n,m} \ . \label{eq:orth}
\end{equation}
For example, if we choose $u=1$, conservation of particles and energy correspond to the simple identities
\begin{eqnarray}
 A_0&=&1 \ \ \ \mathrm{[conservation\ of\ particles]} \label{eq:particles} \\
 A_1&=&0 \ \ \ \mathrm{[conservation\ of\ energy]}    \label{eq:energy}
\end{eqnarray}
provided we make the choice $\alpha=1/2$, as we have in (\ref{eq:expansion}). Because the QLB equation conserves particles and energy, the time derivatives of these two coefficients are identically zero, so if these identities hold for the initial distribution then they hold for all times. The temperature that appears in the expansion is that of the final equilibrated state, which can easily be related to the total (kinetic) energy. The Lenard-Balescu equation also conserves momentum but this is identically zero when we have isotropy in velocity.

Clearly, the expansion (\ref{eq:expansion}) with $u=1$ has many advantages. However, we do pay some price for them. The Laguerre polynomials are orthogonal with respect to the weight $w(x)=x^{1/2} e^{-x}$ and for a function $f(x)$ to be representable by a series of these polynomials it must be square integrable with respect to this weight, i.e.,
\begin{equation}
 \int_0^{\infty} x^{1/2} e^{-x} |f(x)|^2 dx < \infty .
\end{equation}
But because we actually have an expansion of the form
\begin{equation}
 f(x)=e^{-x} \sum_n A_n \Lnhalf(x) ,
\end{equation}
we have the more stringent requirement that $e^x f(x)$ be square integrable, or
\begin{equation}
 \int_0^{\infty} x^{1/2} e^{x} |f(x)|^2 dx < \infty . \label{eq:sqint}
\end{equation}
Say, for example, $f(x)=e^{-x/\gamma}$, then the integral (\ref{eq:sqint}) is
\begin{equation}
 \int_0^{\infty} x^{1/2} e^{x(1-2/\gamma)} dx
\end{equation}
which converges only when $0 < \gamma<2$. For the purposes of this work, the requirement that distributions fall off faster than $e^{-x/2}$ is not particularly problematic. We consider only equilibration problems, in which the end state is the Maxwell distribution, $A_n=\delta_{n0}$, and thus if the initial distribution can be represented then the subsequent evolution can as well. To be more precise, if (\ref{eq:sqint}) is satisfied for the initial time, then it is satisfied for all times. We will not prove this, but it seems very unlikely that the integral in (\ref{eq:sqint}) would be initially finite but then diverge as the distribution becomes more Maxwellian (it is, of course, finite for the Maxwell distribution itself). We will have more to say about this in section \ref{sec:generalizations}, where we show that choosing $u=2$ in (\ref{eq:expansion}) restores completeness at the expense of complicating the collision integrals and conservation conditions. 

To solve the equation, we truncate the expansion (\ref{eq:expansion}) at some $n_\mathrm{max}$, which will be as large as 40 in the present work. The ordinary differential equations that result are of the form
\begin{equation}
 \frac{d A_n}{dt}=\sum_{l=0}^{\nmax} \sum_{k=0}^{\nmax} C^n_{lk}(\{A\}) A_l A_k . \label{eq:ODEs}
\end{equation}
The coefficients $C^n_{lk}$ are integrals over the dielectric function and depend on all the $A_n$, which we denote $\{A\}$, and therefore must be computed on the fly. We describe in section \ref{sec:coeff} how we evaluate these coefficients, but first we turn to the dielectric function.

\section{Dielectric function} \label{sec:dielectric}
It is convenient to define the dimensionless variables
\begin{eqnarray}
 X^2 &\equiv& \frac{\hbar^2 \beta k^2}{4 m_e}  \label{eq:X} \\
 Y^2 &\equiv& \frac{m_e \beta \omega^2}{k^2} , \label{eq:Y}
\end{eqnarray}
in terms of which we will write all of our results. The non-equilibrium dielectric function is derived in Appendix A. In terms of $X$ and $Y$ it is 
\begin{equation}
 \epsilon(X,Y) = 1+\frac{\eta^2}{X^3} w^Q(X,Y) \label{eq:epsquant}
\end{equation}
where $w^Q(X,Y)=w^Q_r(X,Y)+i w^Q_i(X,Y)$ is a complex function whose real and imaginary parts are given by
\begin{eqnarray}
w^Q_r(X,Y) &=& \frac{1}{\sqrt{2}} \sum_{k=0}^\infty A_k \left[-Y_-M \left(k+1,\frac{3}{2};-Y_-^2\right) \right. \cr
&+& \left. Y_+ M \left(k+1,\frac{3}{2};-Y_+^2\right) \right] \\
w^Q_i(X,Y) &=& \sqrt{\frac{\pi}{2}} \frac{1}{2} \sum_k A_k \left[ e^{-Y_-^2} \Lminhalf_k \left( Y_-^2 \right) \right. \cr
&-&\left. e^{-Y_+^2} \Lminhalf_k \left( Y_+^2 \right) \right] 
\end{eqnarray}
where $M(a,b;z)$ is the confluent hypergeometric function, $Y_\pm \equiv (Y \pm X)/\sqrt{2}$, and 
\begin{equation}
\eta \equiv \lambda_Q/\lambda_D
\end{equation}
with
\begin{eqnarray}
 \lambda_Q^2 &\equiv& \frac{\hbar^2 \beta}{4 m_e} \label{eq:lambdaQ} \\
 \lambda_D^2 &\equiv& \frac{1}{4 \pi e^2 n_e \beta} \ .
\end{eqnarray}
Thus, $\eta$ is the ratio of the equilibrium thermal de Broglie and Debye wavelengths. The inverse of this ratio is usually denoted $\Lambda=1/\eta$. At weak coupling, which is where the QLB equation is accurate, $\eta \ll 1$. We will exploit the smallness of $\eta$ when we compute the coefficients. 

Because it greatly simplifies the analysis without detracting from the important physics, we take the limit $\hbar \rightarrow 0$ in the dielectric function. This is equivalent to expanding (\ref{eq:epsquant}) in $X$,
\begin{equation}
 w^Q(X,Y) \approx w^Q(0,Y)+X w(Y),
\end{equation}
where
\begin{equation}
 w(Y) \equiv \left. \frac{\partial w^Q(X,Y)}{\partial X} \right|_{X=0} \ \label{eq:wdef} \ .
\end{equation}
Clearly, $w^Q(0,Y)=0$ and we can compute $w(Y)$ from (\ref{eq:wdef}) by making use of the hypergeometric contiguous relation
\begin{equation}
 z \frac{\partial M(a,b;z)}{\partial z}=(b-1)[M(a,b-1;z)-M(a,b;z)].
\end{equation}
The dielectric function is then
\begin{equation}
 \epsilon_\mathrm{cl}(X,Y)=1+\frac{\eta^2}{X^2} w(Y) \label{eq:eps}
\end{equation}
with the real and imaginary parts of $w(Y)$ given by
\begin{eqnarray}
 w_r(Y)&=&\sum_{k=0}^\infty A_k M \left(k+1,\frac{1}{2};-\frac{Y^2}{2} \right) \label{eq:wr} \\
 w_i(Y)&=& \sqrt{\frac{\pi}{2}} Y e^{-\frac{Y^2}{2}} \sum_{k=0}^\infty A_k L_k^{(1/2)} \left(\frac{Y^2}{2}\right) . \label{eq:wi}
\end{eqnarray}
Neglecting quantum effects in the dielectric function probably does not impact the solution in a major way and, of course, we retain this physics everywhere else in the QLB equation. With only a few tens of parameters, namely the $A_k$, to be determined numerically, this analytic representation of the dielectric function is very convenient. We can, for example, use it to determine the dispersion relation of waves in non-equilibrium plasmas. We will see how this form is also useful in the numerical solution of the QLB equation, despite the presence of the confluent hypergeometric function.

Note that we can, if we wish, simplify the problem even further by considering only static screening and setting $Y=0$ in the dielectric function. The result is
\begin{equation}
 \epsilon_\mathrm{static}(X)=1+\frac{\eta^2}{X^2} \sum_{k=0}^\infty A_k . \label{eq:static}
\end{equation}
In this form, we no longer have dynamical screening effects but the static screening length is still calculated from the distribution.

\section{Coefficients} \label{sec:coeff}
The coefficients of equation (\ref{eq:ODEs}) are computed by multiplying the equation by a Laguerre polynomial and integrating over velocity. The details of this are given in Appendix B. The result is
\begin{eqnarray}
C^n_{lk} &=& - C_0 \frac{n!}{\Gamma(n+3/2)} \cr 
& & \int_0^{\infty} dX \frac{e^{-X^2}}{X^3} \int_{-\infty}^{\infty} dY \frac{e^{-Y^2}}{|\epsilon(X,Y)|^2} P^n_{lk}(X,Y) \label{eq:Cnlk}
\end{eqnarray}
where the prefactor is
\begin{equation}
 C_0 \equiv \frac{n_e \beta^{3/2} \sqrt{\pi} e^4}{\sqrt{m_e}} \ .
\end{equation}
The functions $P^n_{lk}(X,Y)$ are polynomials in $X$ and $Y$ defined by,
\begin{equation}
 P^n_{lk}(X,Y) = \frac{1}{2}[q^n_{lk}(X,Y)+q^n_{lk}(X,-Y)]
\end{equation}
where
\begin{eqnarray}
 & &q^n_{lk}(X,Y) \equiv \sum_{j=0}^{\min(l,n)}\Lminhalf_{n-j} \left(Y_-^2 \right) \cr
 &\times& \left[ \Lminhalf_k \left(Y_+^2 \right) \Lminhalf_{l-j}\left(Y_-^2 \right) \right. \cr 
 & &\ \ \ \ \ \ \ \ \ \  -\left. \Lminhalf_k \left(Y_-^2 \right) \Lminhalf_{l-j} \left( Y_+^2 \right) \right] \label{eq:q}
\end{eqnarray}
so $P$ is just the even part of $q$ in $Y$. It is not really necessary to take the even part explicitly because the integration over $Y$ filters out the odd powers, but we do it to make the following analysis more clear. Note also that the symmetry of $q$ in $X$ and $Y$ means that $P^n_{lk}$ contains only even powers of $X$ and it turns out that $X^2 Y^2$ is the lowest power for all $n,l,k$. To facilitate our approximations, we use the decomposition
\begin{equation}
 P^n_{lk}(X,Y) = X^2 Y^2 G^n_{lk}(Y) + R^n_{lk}(X,Y) \label{eq:decomp}
\end{equation}
where $G^n_{lk}(Y)$ is a polynomial in $Y$ and $R^n_{lk}(X,Y)$ contains terms only of order $X^4$ and higher. Now, the $X$ integration in (\ref{eq:Cnlk}) would of course be divergent as $X \rightarrow 0$ were it not for the dielectric function. However, only the first term in (\ref{eq:decomp}) would actually diverge. Our decomposition is therefore a separation into the term that needs the dielectric function for convergence, and the rest of the integrand that does not. From here on, we will keep the dielectric function only where it is actually needed for convergence and set it to $1$ elsewhere. This approximation can be justified as follows. 

Physical parameters, such as mass and density enter into the coefficients in (\ref{eq:Cnlk}), and therefore the equation, in two places: the prefactor $C_0$ and the dimensionless ratio $\eta$. The constant $C_0$ only sets the overall time scale of the problem and two solutions with the same $\eta$ and initial distribution but different $C_0$ will be identical up to time rescaling. Therefore, the only really important quantity is $\eta$, and varying things like the particle mass, the number density and the final equilibrium temperature only matters to the extent that we are changing $\eta$. As such, as previously mentioned, we stick with electrons at $10^{25}\mathrm{cm}^{-3}$. The latter choice makes $\eta$ similar with the coupling constant, $\Gamma$, as we vary the temperature. As we will show later, the expansion of $C^n_{lk}$ in $\eta$ is
\begin{equation}
 C^n_{lk}=a_0 + \sum_{i=0}^\infty b_i \eta^{2 i} \ln \eta + \sum_{i=1}^\infty a_i \eta^{2i} \ .
\end{equation}
The term proportional to $b_0$ is the Coulomb logarithm and $a_0$ is the order-unity term; if $\eta$ is small, we may be justified in neglecting the rest. And if $\eta$ is smaller still, the logarithmic term will dominate $a_0$ and the Landau equation is fine. However, for arbitrary non-equilibrium initial conditions, the $\mathrm{O}(1)$ terms depend on the distribution and must be computed before we can be sure they can be neglected, making the definition of ``small'' for $\eta$ highly problem-dependent. Our strategy of keeping the dielectric function only where it is necessary for convergence is equivalent with computing $a_0$ and $b_0$ and dropping the rest. Thus, we neglect terms that are $\mathrm{O}(\eta^2 \log \eta)$ and higher, which does not make a great difference in many cases. For example, at a density of $10^{25} \mathrm{cm}^{-3}$ at $T=1000$ eV, $\eta \approx 0.05$ and $\eta^2 \ln \eta \approx -7.5 \times 10^{-3}$, compared with the term $\ln \eta \approx -3$ and other $\mathrm{O}(1)$ terms that we are going to keep. We discuss below some situations where one might need the higher-order terms, but we will not be concerned about computing them in this paper. In any case, it is a straightforward generalization to include them (see section \ref{sec:generalizations}) but, of course, this becomes more computationally expensive. 

Under this approximation, the integrals over $R^n_{lk}$, as they do not contain the $A_k$, can be precomputed. We define the coefficients
\begin{equation}
 B^n_{lk} \equiv \frac{n!}{\Gamma(n+3/2)} \int_0^{\infty} dX \frac{e^{-X^2}}{X^3} \int_{-\infty}^{\infty} dY e^{-Y^2} R^n_{lk}(X,Y) . \label{eq:Bnlk}
\end{equation}
Another set we will need is
\begin{equation}
 S^n_{lk} \equiv \frac{n!}{\Gamma(n+3/2)} \int_{-\infty}^{\infty} dY e^{-Y^2} Y^2 G^n_{lk}(Y) \ .
\end{equation}
Even though these coefficients can be precomputed, doing so is not completely trivial. As $n$, $l$ and $k$ become large, $R^n_{lk}$ becomes higher-order in $X$ and $Y$. For example, at $(n,l,k)=(40,40,40)$, the most difficult case, $R^n_{lk}(X,Y)$ is order 234 in $X$ and $Y$. If we wish to use a numerical integration scheme for this we need to evaluate $R^n_{lk}(X,Y)$ at the quadrature points, which can prove to be tricky with such high order polynomials. There is probably an optimal solution to this problem, but we resort to brute force. We use the CLN arbitrary precision library\cite{CLN} for C++ and we decompose the polynomial $R^n_{lk}(X,Y)$ into its powers,
\begin{equation}
 R^n_{lk}(X,Y) = \sum_{ij} a^{n}_{lkij} X^{2i} Y^{2j} \ , \label{eq:Rdecomp}
\end{equation}
where the sums over $i$ and $j$ start at $i=j=2$. Using the exact integrals
\begin{equation}
 \int_0^\infty e^{-X^2} X^{2i-3} dX = \frac{1}{2} \Gamma(i-1)
\end{equation}
and
\begin{equation}
 \int_{-\infty}^\infty e^{-Y^2} Y^{2j} dY = \Gamma \left( \frac{1}{2}+j \right) \ ,
\end{equation}
where $\Gamma(x)$ is the gamma function, combined with the decomposition (\ref{eq:Rdecomp}) allows us to evaluate (\ref{eq:Bnlk}) so long as we have sufficient precision; we keep 60 digits for this purpose. Of course, we do not need this many when we solve the actual differential equation, so in the end we keep the resulting $B^n_{lk}$ only to double precision. The constants $S^n_{lk}$ can be handled in the same way but in equation (\ref{eq:exactSnlk}) we give the exact solution for these.

Now we are left with the problem of evaluating
\begin{equation}
 I^n_{lk} \equiv \int_0^{\infty} dX \frac{e^{-X^2}}{X} \int_{-\infty}^{\infty} dY \frac{e^{-Y^2}}{|\epsilon(X,Y)|^2} Y^2 G^n_{lk}(Y) \ , \label{eq:Fnlk}
\end{equation}
which must be computed on the fly. The strategy is to compute the $X$ integral exactly, which would hardly be possible if we were not using the classical dielectric function. The remaining one-dimensional integral over $Y$ will contain a tangle of special functions, but the integrand is smooth and can easily be handled with straightforward Gaussian quadrature. The steps required to reduce (\ref{eq:Fnlk}) are given in Appendix C, with the result,
\begin{equation}
 I^n_{lk}=-\frac{\Gamma(n+3/2)}{n!} \left[S^n_{lk}\left( \frac{\gamma_E}{2}+ \ln \eta \right) + F^n_{lk} \right] \label{eq:Inlk}
\end{equation}
where
\begin{equation}
 F^n_{lk} \equiv \frac{n!}{\Gamma(n+3/2)} \frac{1}{2} \int_{-\infty}^\infty dY e^{-Y^2} Y^2 G^n_{lk}(Y) F(Y) ,
\end{equation}
and
\begin{eqnarray}
 F(Y) &\equiv& \frac{1}{2} \ln [w_r^2(Y)+w_i^2(Y)] \cr 
 & & \ \ \ \ \ \ \ \ \ \ + \frac{w_r(Y)}{w_i(Y)} \arctan [w_r(Y),w_i(Y)] \label{eq:FY}
\end{eqnarray}
where $\arctan(x,y)$ is the quadrant-correct version of $\tan^{-1} y/x$, producing an angle in the range $(-\pi,\pi]$. The integrand in (\ref{eq:Fnlk}) is well-behaved, without any of the sharp peaks that typically characterize dielectric function integrands, and we avoid the need for any pole-correcting integration techniques \cite{Vorberger, Chapman}. The same strategy was used by Williams and DeWitt \cite{Williams} for conductivity calculations in a two-component plasma in equilibrium . Although a very different problem from ours, it involves the same collision operator and the same kinds of integrals (compare their equation (73) with our (\ref{eq:FY})). In fact, this method would be useful for other problems as well, such as temperature equilibration \cite{Daligault, Benedict}.

Although we now have a one-dimensional integral, we are still faced with the task of evaluating it at every time step. The factor $e^{-Y^2}$ in the integrand strongly suggests we use Gauss-Hermite quadrature. Actually, because the integrand is even, we make the substitution $x=Y^2$ and use a closely related Gauss-Laguerre scheme. We then have
\begin{eqnarray}
& & \int_{-\infty}^\infty dY e^{-Y^2} Y^2 G^n_{lk}(Y) F(Y) \cr
 & & \ \ \ \ =\int_0^\infty dx x^{1/2} e^{-x} G^n_{lk}(\sqrt{x}) F(\sqrt{x}) \cr 
 & & \ \ \ \ \approx \sum_{i=1}^{N_p} W_i G^n_{lk}(\sqrt{x_i}) F(\sqrt{x_i})
\end{eqnarray}
where $x_i$ are the abscissa points, the zeros of $L_{N_p}^{(1/2)}(x)$, and $W_i$ are the weights, given by
\begin{equation}
 W_j=\frac{x_j \Gamma(N_p+1/2)}{N_p! (N_p+1/2) \left[L_{N_p-1}^{(1/2)}(x_j)\right]^2}  \ . \label{eq:quad}
\end{equation}
We choose $N_p=200$ to ensure that we have an accurate integration even for the largest $n$, $l$, and $k$. This number can probably be varied to optimize performance, and it may not always be necessary to include every term in (\ref{eq:quad}), especially for the smaller $(n,l,k)$. We do not explore this particular performance issue too closely, but, as we will show, this scheme is more than sufficiently accurate for our purposes and enables a numerical solution of the Lenard-Balescu equation. What is more, because our integration is now simply a sum over quadrature points, the values of the confluent hypergeometric functions and Laguerre polynomials that appear in $w_r(\sqrt{x})$ and $w_i(\sqrt{x})$ can be precomputed at $x_i$ and never need to be evaluated during the solution. We will also precalculate the points $G^n_{lk}(\sqrt{x_i})$. This is somewhat tricky because, exactly like $R^n_{lk}$, as $(n,l,k)$ become large, the order of this polynomial also becomes large and we require high precision to evaluate it. To do this, we use a strategy similar to the one we employed to calculate $B^n_{lk}$. First, we compute the quadrature points, $x_i$, to high precision using Mathematica. Then we decompose $G^n_{lk}(Y)$ as in (\ref{eq:Rdecomp}) and evaluate each power of $Y$ at $\sqrt{x_i}$ to 60 digits and sum these results to get $G^n_{lk}(\sqrt{x_i})$. As before, we keep these values only to double precision, so no arbitrary precision library is needed in the actual solver. 

\begin{figure}
\begin{center}
\includegraphics[width=3in]{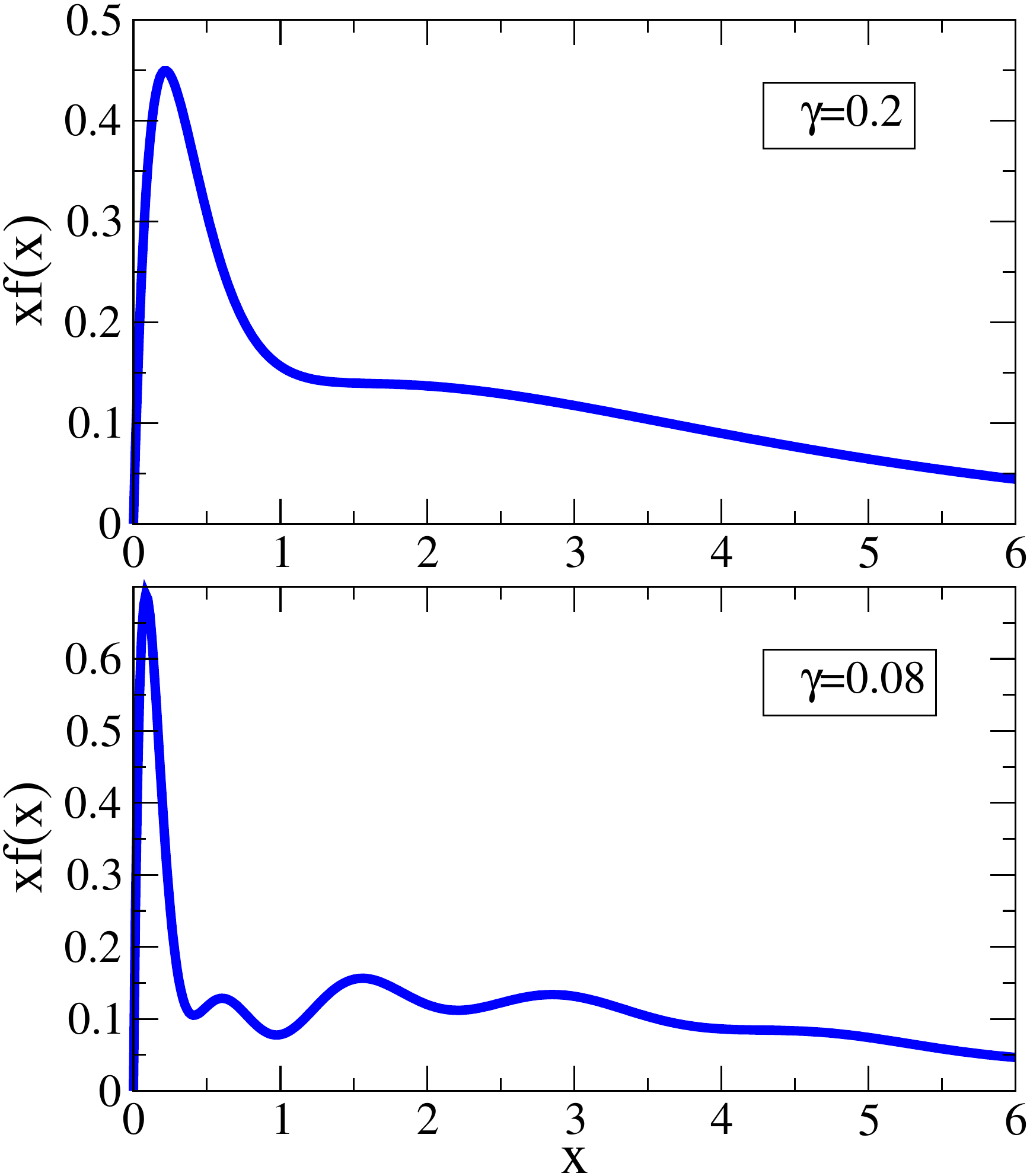}
\caption{Two-temperature initial distribution using $\nmax=40$ polynomials with $\gamma=0.2$ (top) and $\gamma=0.08$ (bottom). The $\gamma=0.08$ case is under-resolved with $\nmax=40$, and the bumps are mostly a result of this, but the distribution is positive everywhere and properly normalized and is therefore suitable as an initial condition. The variable $x=\beta m_e v^2/2$ and $f(x)$ is divided by $n_e (\beta m_e/2 \pi)^{3/2}$.}
\label{fig:twoTinit}
\end{center}
\end{figure}

To demonstrate the accuracy of the quadrature scheme, we will compute $F^n_{lk}$ for $(n,l,k)=(40,40,40)$, the most difficult case. Of course, we must also specify a distribution and for this we use a two-temperature plasma in which half the particles are at temperature $T_1$ and the other half at $T_2$, which is shown in the top panel of Figure \ref{fig:twoTinit}. The coefficients for this distribution are calculated in section \ref{sec:IC} and are characterized by the single parameter $\gamma \equiv T_1/T$, where $T$ is the final temperature, which we choose to be $0.2$, the limit of our resolution ability. Using Mathematica's adaptive numerical integration with 100-digit precision, we find that
\begin{equation}
\int_{-\infty}^\infty dY e^{-Y^2} Y^2 G^{40}_{40,40}(Y) F(Y) \approx 2.1563096073 \ . \label{eq:FYint}
\end{equation}
The calculation takes several minutes but is accurate to the number of digits presented. For this integral, our double precision quadrature code gives $2.1563096082$, correct to eight decimal places, far more than we need, and is just a sum over 200 points. Of course, the accuracy depends on the distribution and we may not always achieve this level. For example, consider $\gamma=0.08$, which is badly under-resolved when $\nmax=40$. However, the Laguerre series is positive everywhere, as shown in the bottom panel of Figure \ref{fig:twoTinit}, and thus is an acceptable distribution. Done with adaptive integration in high precision, the integral (\ref{eq:FYint}) is $2.9349261394$. With our Gaussian quadrature scheme we find $2.9323983879$, which is not disastrous but not nearly as accurate as in the previous example, probably due to the oscillations in the distribution. Adding more quadrature points would probably improve the accuracy, but computing the integral to three figures is sufficient for our purposes.

We have a fast and accurate method for integrating over the dielectric function but the price we pay for this is that we must keep a huge number of precalculated values; if we want to use $\nmax=40$ coefficients in the polynomial expansion and $N_p=200$ quadrature points, the file containing the $G^n_{lk}(\sqrt{x_i})$ is 287 megabytes. While this is manageable enough, when we consider that the number of coeffients needed grows as $\nmax^3$ and that we will need more quadrature points as we increase $(n,l,k)$, it is clear that this can quickly grow out of control. However, the present approach is surely the brute-force method to compute $C^n_{lk}$, and there are likely better ways to do this. For example, let us define $V^j_{lk}$ to be the integral over triple products of Laguerre polynomials,
\begin{eqnarray}
 V^j_{lk} &\equiv& \int_{-\infty}^\infty dY \int_0^\infty dX \frac{e^{-Y^2}}{|\epsilon(X,Y)|^2} \frac{e^{-X^2}}{X^3} \times \cr
& &L_j^{(-1/2)}\left(Y_-^2 \right) L_k^{(-1/2)}\left(Y_+^2 \right) L_l^{(-1/2)}\left(Y_-^2 \right) \label{eq:V} \ .
\end{eqnarray}
The equation (\ref{eq:Cnlk}) for the coefficients can then be written
\begin{equation}
 C^n_{lk} = -C_0 \frac{n!}{\Gamma(n+3/2)} \sum_{j=0}^{\mathrm{min}(l,n)} (V^{n-j}_{l-j,k} - V^{n-j}_{k,l-j}) \ .
\end{equation}
The $V^j_{lk}$ satisfy a recurrence formula that can possibly be exploited to facilitate computation of $C^n_{lk}$ without needing huge files of precomputed data. In Appendix E, we derive this formula and show that it has an exact solution. It may well be that such a method is superior once certain mathematical issues are resolved.

Putting the pieces of the present method together, we find for the coefficients
\begin{equation}
 C^n_{lk} = C_0 \left[ S^n_{lk} \left(\frac{\gamma_E}{2} + \ln \eta \right) - B^n_{lk} + F^n_{lk}(\{A\}) \right] \label{eq:Cfinal}
\end{equation}
which, as mentioned, neglects terms that are $\mathrm{O}(\eta^2 \ln \eta)$. However, we have derived the Coulomb logarithm, rather than imposing it, along with all $\mathrm{O}(1)$ terms that arise from the quantum Lenard-Balescu equation. These are by far the dominant contributions to the equation for the situations we will consider. On the other hand, as we can see from equation (\ref{eq:exactX}) in Appendix C, the expansion of the incomplete gamma function that leads to (\ref{eq:Cfinal}) is not in $\eta^2$ but in $\eta^2 w(Y)$. It is conceivable that some distributions might make $|w(Y)|$ comparable to $\eta^{-2}$ and then we would not be justified in discarding these terms. For example, if we have a two-temperature initial condition we can make $|w(0)|$ as large as we want by increasing the temperature separation. We will not encounter such an extreme situation here but, as mentioned, we discuss how to restore these terms in section \ref{sec:generalizations}. 
\section{Special Cases} \label{sec:special}
By neglecting different terms in (\ref{eq:Cfinal}), we can find solutions to various kinetic equations. These are listed here.

\subsection{Landau equation}
If we drop the quantum diffraction terms $B^n_{lk}$ and the screening terms $F^n_{lk}$, we are left with the coefficients of the Landau equation,
\begin{eqnarray}
 C^n_{lk} &=& C_0 S^n_{lk} \left(\frac{\gamma_E}{2} + \ln \eta \right) \\
 &=& -C_0 S^n_{lk} \log \Lambda \ . \label{eq:LandauCnlk}
\end{eqnarray}
In the first line we write the Coulomb logarithm informed by the Lenard-Balescu equation, but one is free to make any choice of $\log \Lambda$ one wishes; the integrals in this equation are divergent and the form of $\Lambda$ results from the choice of cutoffs. The coefficients $C^n_{lk}$ are all precomputed so nothing needs to be calculated on the fly. This makes the solution extremely cheap compared with the full QLB equation. Another fact worth mentioning is that the coefficients $S^n_{lk}$ actually have a closed form in terms of hypergeometric functions,
\begin{eqnarray}
 & & S^n_{lk}= \frac{\sqrt{\pi}}{2 \sqrt{2}} \frac{n!}{\Gamma(n+3/2)}\frac{1}{4^k} \frac{1}{n! l! k!} \Gamma \left(\frac{3}{2}-n+k+l \right) \times \cr
 & & \Gamma \left(-\frac{1}{2}+n+k-l \right) \left \{ 16\ _3 \widetilde{F}_2 \left(1,-l,n;b_1^1,b_2^1;1 \right) \right. \cr
 &-& \left. [(2k-2l+2n)^2-1]\ _3 \widetilde{F}_2 \left(1,-l,n;b_1^2,b_2^2;1 \right) \right\} \label{eq:exactSnlk}
\end{eqnarray}
where 
\begin{eqnarray}
 b_1^1&=&-\frac{1}{4}+\frac{k-l-n}{2} \\
 b_2^1&=&\frac{1}{4}+\frac{k-l-n}{2} \\
 b_1^2&=&\frac{3}{4}+\frac{k-l-n}{2} \\
 b_2^2&=&\frac{5}{4}+\frac{k-l-n}{2}
\end{eqnarray}
and $_3 \widetilde{F}_2(a_1,a_2,a_3;b_1,b_2;z)$ is a regularized hypergeometric function. The latter is defined by
\begin{equation}
 _3 \widetilde{F}_2(a_1,a_2,a_3;b_1,b_2;z) \equiv \frac{_3F_2(a_1,a_2,a_3;b_1,b_2;z)}{\Gamma(b_1) \Gamma(b_2)}
\end{equation}
where $_pF_q(a_1...,a_p;b_1,...,b_q;z)$ is the generalized hypergeometric function. These expressions may not seem terribly convenient. However, Mathematica, and probably other similar programs, quickly evaluates them and easily handles the differential equations too. This prescription therefore provides a fast and convenient way to solve the single-component Landau equation. The derivation of (\ref{eq:exactSnlk}) is given in Appendix D.

\subsection{Non-degenerate quantum Landau equation}
Setting $F^n_{lk}=0$ in (\ref{eq:Cfinal}) neglects the dielectric function but we still have the quantum wave effects embodied in the $B^n_{lk}$ and we end up with the coefficients for what we call the non-degenerate quantum Landau equation. The reason for this ungainly term is that ``quantum Landau equation'' is already in use \cite{Hu,Daligault2016} for a kinetic equation that accounts for quantum statistics but no other quantum effects, which is sort of the complement of our equation. The integrals in this equation are divergent as $X \rightarrow 0$ but converge as $X \rightarrow \infty$, meaning that we need only a lower cutoff. This will, of course, generally be chosen to be the equilibrium Debye length, or $X_c=\eta$ in our dimensionless variables. The resulting coeffients are
\begin{equation}
 C^n_{lk} = C_0 \left[ S^n_{lk} \left(\frac{\gamma_E}{2} + \ln \eta \right) - B^n_{lk} \right] \ ,
\end{equation}
which can once again all be precalculated.

\subsection{Classical Lenard-Balescu equation}
Finally, we can set $B^n_{lk}=0$ but retain $F^n_{lk}$. The coefficients are then
\begin{equation}
 C^n_{lk} = C_0 \left[ S^n_{lk} \left(\frac{\gamma_E}{2} + \ln \eta \right) + F^n_{lk}(\{A\}) \right] \ .
\end{equation}
These correspond to the classical Lenard-Balescu equation, in which we cure the $X \rightarrow \infty$ divergence by introducing a cutoff in wavenumber at the inverse of the thermal deBroglie wavelength, $\lambda_Q$ in equation (\ref{eq:lambdaQ}), or $X_c=1$ in the dimensionless units. Of course, one can instead cut the integral off at the Landau length to keep everything classical.

\subsection{Quantum Lenard-Balescu with static screening}
The dielectric function that neglects dynamical screening is given in equation (\ref{eq:static}). This corresponds to the choices,
\begin{eqnarray}
 w_r(Y)&=& \sum_{k=0}^\infty A_k \cr
 w_i(Y)&=& 0 \ .
\end{eqnarray}
Using these in equation (\ref{eq:FY}), we find
\begin{equation}
 F(Y)=\ln \sum_{k=0}^\infty A_k + 1
\end{equation}
leading to the coefficients
\begin{equation}
 C^n_{lk} = C_0 \left\{ \frac{1}{2} S^n_{lk} \left[\gamma_E + 1 + \ln \left( \eta^2 \sum A_k \right) \right] - B^n_{lk} \right\}.
\end{equation}
We can see that static screening contributes an additional constant (i.e., one) and modifies the Coulomb logarithm by the sum over $A_k$. This provides a correction to the Debye length 
and is trivial to compute.

\section{Initial conditions} \label{sec:IC}
To test our algorithm, we consider the relaxation to equilibrium of various initial distributions. For general $u$, the coefficients $A_n(0)$ for a given $f(v,0)$ are 
\begin{eqnarray}
 & & A_n(0)=\frac{2 \pi^{3/2} u^{3/2}}{n_e}\frac{n!}{\Gamma(n+3/2)} \cr 
 &\times& \int_0^\infty e^{-\frac{\beta m v^2}{2}(u-1)}v^2 f(v,0) \Lnhalf\left(u \betavsq\right) dv, \label{eq:initialA}
\end{eqnarray}
which is an easy consequence of the orthogonality property of Laguerre polynomials.

\subsection{Two-temperature plasma}
Here we will consider the case of a two-temperature one-component plasma. A number density $n_1$ have temperature $T_1$ and $n_2$ have $T_2$ so,
\begin{eqnarray}
 f(v,0)&=&\left(\frac{m_e}{2 \pi}\right)^{\frac{3}{2}} \left[n_1 \beta_1^{3/2} \exp \left(-\frac{m_e \beta_1 v^2}{2}\right) \right. \cr
 &+& n_2 \left. \beta_2^{3/2} \exp \left(-\frac{m_e \beta_2 v^2}{2}\right) \right] \ .
\end{eqnarray}
We define the fractions $\xi \equiv n_1/n_e$, $\xi_2 \equiv n_2/n_e$, $\gamma \equiv T_1/T$, and $\gamma_2 \equiv T_2/T$. By conservation of particles and energy we have
\begin{eqnarray}
 \xi_2&=&1-\xi \\
 \gamma_2 &=& \frac{1-\xi \gamma}{1-\xi} .
\end{eqnarray}
We make the choice $\gamma<1$, so that $\gamma_2>1$. Carrying out the integration (\ref{eq:initialA}) we find,
\begin{equation}
 A_n(0)= \frac{\xi(1-\gamma)^n u^{3/2}}{(\gamma u - \gamma +1)^{n+3/2}} + \frac{(1-\xi)(1-\gamma_2)^n u^{3/2}}{(\gamma_2 u - \gamma_2+1)^{n+3/2}} \ . \label{eq:twoT}
\end{equation}
For $u=1$, which is what we use exclusively here, the condition (\ref{eq:sqint}) means that $\gamma_2$ must be less than 2 or the expansion does not converge, which is also clear enough in (\ref{eq:twoT}). This is, of course, a purely mathematical requirement and it leads to the constraint between $\xi$ and $\gamma$,
\begin{equation}
 \gamma > \frac{2 \xi-1}{\xi} \ . \label{eq:constraint}
\end{equation}
If we needed to break this we would choose a different value of $u$, such as 2. However, (\ref{eq:constraint}) is no constraint if $\xi \leqslant 1/2$ so if, for example, we have an equal number of particles of each temperature then $\gamma$ can be chosen arbitrarily in the range $[0,1]$ with $\gamma_2<2$ enforced by conservation of energy. This is the first situation we will consider.

\begin{figure}
\begin{center}
\includegraphics[width=3.25in]{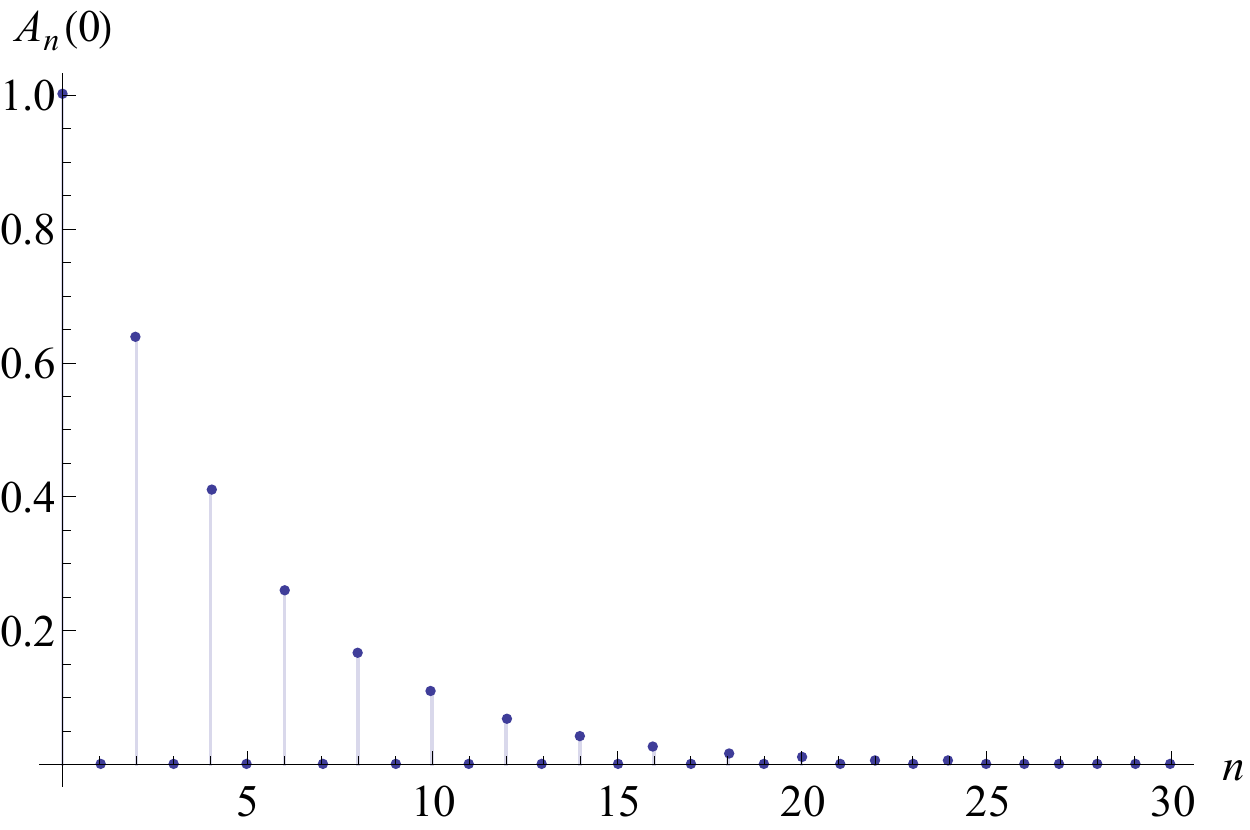}
\caption{The coefficients, $A_n(0)$, for the two-temperature initial condition with $\gamma=0.2$.}
\label{fig:twoTcoeffs}
\end{center}
\end{figure}

\begin{figure}
\begin{center}
\includegraphics[width=3.25in]{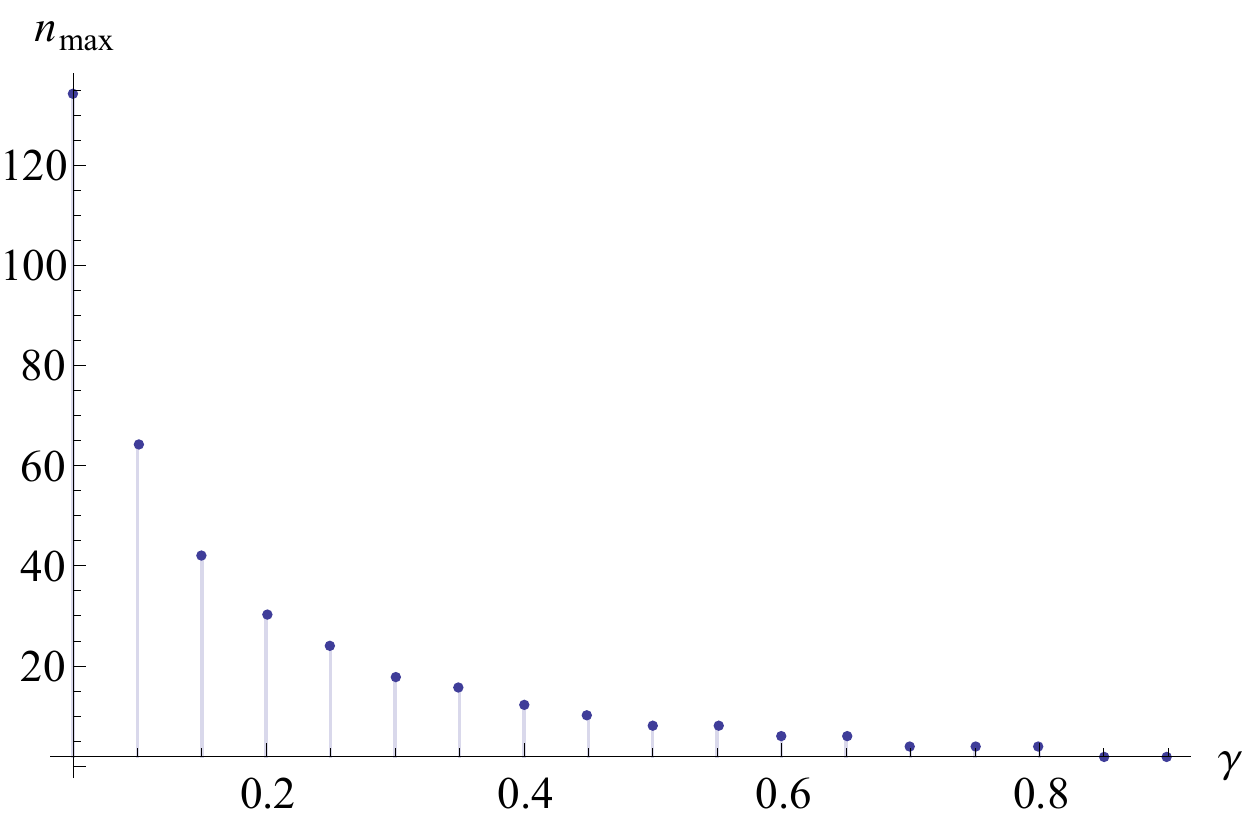}
\caption{The number of polynomials needed such that $A_{\nmax}<10^{-3}$ as a function of $\gamma$ for a two-temperature plasma. The left-most point corresponds to $\gamma=0.01$ \ . In this paper, we have $\nmax=40$ and thus we begin to have resolution problems when $\gamma<0.2$ \ .}
\label{fig:malpha}
\end{center}
\end{figure}

With $\xi=1/2$, the coefficients $A_n(0)$ for $\gamma=0.2$ and $u=1$ are shown in Figure \ref{fig:twoTcoeffs}. It is a mathematical peculiarity that, for this situation, every odd coefficient is zero. In an equilibration problem, the distribution becomes more Maxwellian with time and thus the initial condition is probably the most difficult thing to resolve. In other words, one will not need more polynomials at a later time than at the beginning. To get an idea of the number needed for a two-temperature system, we consider the $\nmax$ at which we first have $A_{\nmax} < \delta$. For this problem,
\begin{equation}
 N=\frac{\log \delta}{\log (1-\gamma)} \ . \label{eq:Ngamma}
\end{equation}
Shown in Figure \ref{fig:malpha} is a plot of $N(\gamma)$ using $\delta = 10^{-3}$. This is a somewhat arbitrary choice, to be sure, but it provides a useful rule of thumb. From this plot it is clear that for $n=40$, which is our maximum, one would not want to go far below $\gamma=0.2$. We can, of course, invert (\ref{eq:Ngamma}) to estimate the minimum $\gamma$ for a given $N$,
\begin{equation}
 \gamma=1-\delta^{1/N} \ .
\end{equation}
We should point out that although the initial $A_n$ depend only on the ratio of the initial to the final temperature, their subsequent values will depend on the absolute temperature through the dependence of the coefficients on $\eta$ and the prefactor.
\subsection{Gaussian distribution} \label{sec:Gauss_init}
We consider here the initial distribution
\begin{equation}
 f(v)=B e^{-\frac{(v-v_0)^2}{2 \sigma^2}} , \label{eq:gaussian}
\end{equation}
for demonstration purposes, not because we have a particular application in mind. The amplitude, $B$, and variance, $\sigma^2$, can be related to the number density and the energy by first defining the integrals
\begin{eqnarray}
 I_1(\bar{v}_0)&\equiv& \int_0^\infty v^2 e^{-(v-\bar{v}_0)^2} dv \label{eq:I1} \\
 I_2(\bar{v}_0)&\equiv& \int_0^\infty v^4 e^{-(v-\bar{v}_0)^2} dv    \label{eq:I2}
\end{eqnarray}
with $\bar{v}_0 \equiv v_0/\sqrt{2} \sigma$. We then have
\begin{eqnarray}
 \sigma^2 &=& \frac{3 I_1(\bar{v}_0)}{2 I_2(\bar{v}_0) \beta m_e} \\
 B &=& \frac{n_e}{4 \pi (\sqrt{2} \sigma)^3 I_1(\bar{v}_0)} \ .
\end{eqnarray}
The integrals (\ref{eq:I1}) and (\ref{eq:I2}) can be expressed in terms special functions, but they are easily evaluated numerically for a given $\bar{v}_0$. This parameter is the only one on which the $A_n$ depend. For general $u$, these are given by
\begin{eqnarray}
& &A_n = \sqrt{\frac{\pi}{2}} \frac{[I_2(\bar{v}_0)]^{3/2}}{3^{3/2} [I_1(\bar{v}_0)]^{5/2}} \frac{n!}{\Gamma(n+3/2)} \cr
 &\times& \int_0^{\infty} e^{-\left(1-\frac{1}{u}\right)x-\left(\sqrt{ 2 I_2(\bar{v}_0)/[3 I_1(\bar{v}_0)u] }x^{1/2}-\bar{v}_0 \right)^2} \cr 
& & \ \ \ \ \ \ \ \ \ \ \  \Lnhalf(x)x^{1/2}dx \ .
\end{eqnarray}
Although the integral can be written exactly in terms of Hermite polynomials using formula 7.374.9 of Gradshteyn and Ryzhik \cite{Gradshteyn} we just solve it numerically. The parameter $\bar{v}_0$ determines the number of polynomials needed to resolve the distribution. Being limited to 40 polynomials, we find that we can choose $\bar{v}_0$ no larger than 2. Exactly what this means in terms of absolute velocity depends on the values of the other parameters, such as $n_e$ and $\beta$.

\section{Differential equations}
The ordinary differential equations do not turn out to be very difficult to solve. We use the fifth-order Runge-Kutta scheme with adaptive time step implemented in the Boost library \cite{Schling}, which easily handles the problem. For the Landau equation, and any of the others for which the coefficients can be precalculated, the solution is found more or less instantaneously using $\nmax=40$ polynomials. For the Lenard-Balescu equation the story is different and a solution can take several hours, but the bulk of the work is in the computation of the $C^n_{lk}$ with the equation itself not being any more difficult than the other cases. This can be easily sped up with parallel computation; each processor computes every $C^n_{lk}$ for a different range of $n$. The results are then shared and the equation can be solved on a single processor. This scheme scales essentially perfectly with the number of processors and in practice we generally assign one $n$ to each processor.

\section{Results} \label{sec:results}

\subsection{Comparison with Fokker-Planck solution}
\begin{figure}
\begin{center}
\includegraphics[width=3.25in]{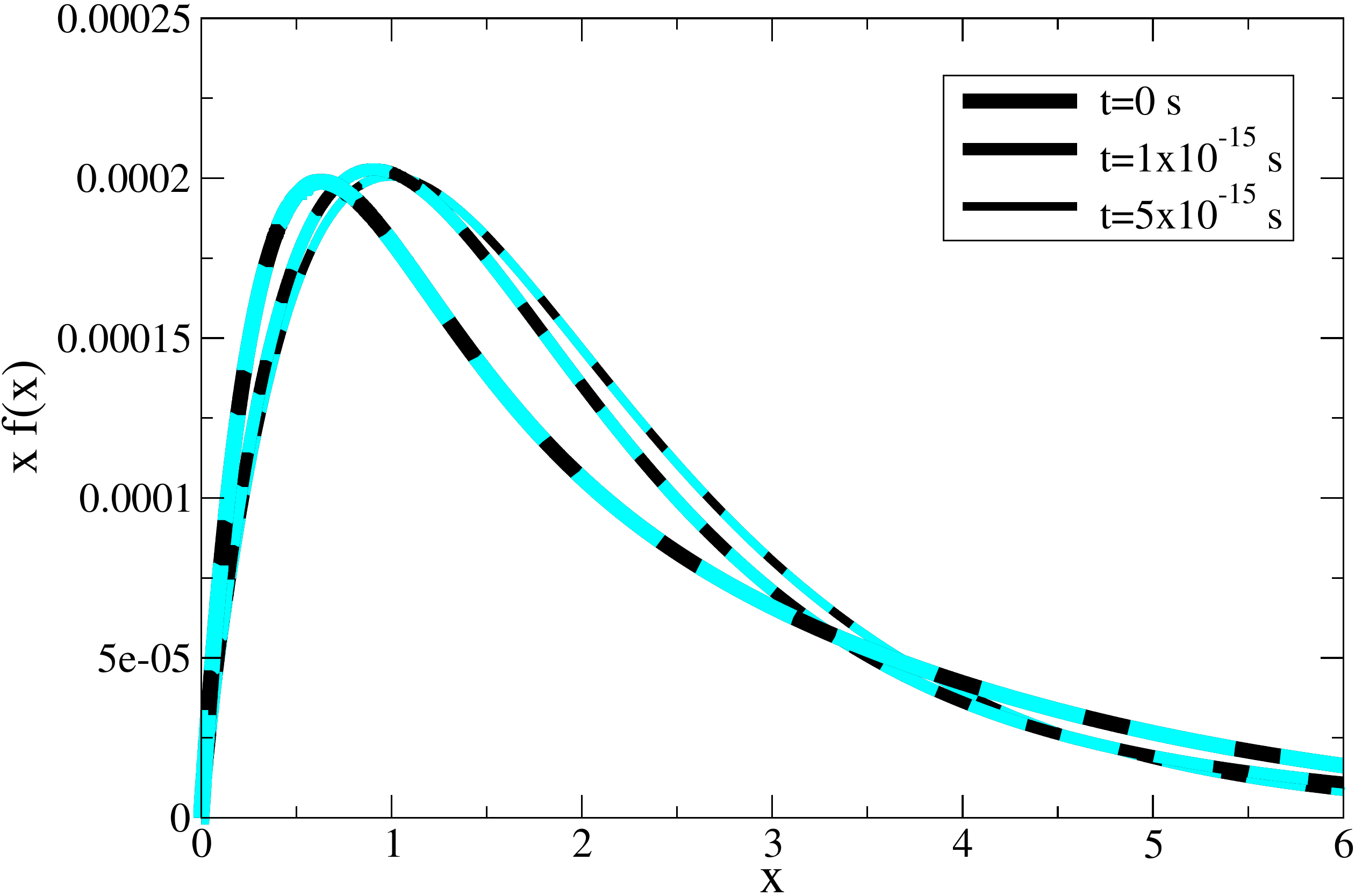}
\caption{Comparison between distributions calculated with our spectral solution of the Landau equation (solid black) and a discretized Fokker-Planck solution (dashed cyan). The two are indistinguishable for all times, three of which are shown.}
\label{fig:FP}
\end{center}
\end{figure}

The first thing we wish to do is check that our approach is sound by comparing our solution to the Landau equation with the result of a more traditional discretized solution to the Fokker-Planck equation. Data for this was provided by David Michta using a code he developed to study thermonuclear burn \cite{Michta}. This approach uses discretization in velocity that is designed to ensure conservation of particles \cite{Chang} and energy \cite{Epperlein}, non-trivial problems in discretization schemes. The situation we considered was a two-temperature one-component plasma of electrons at a density of $2 \times 10^{25} \mathrm{cm}^{-3}$. Half the particles are Maxwellian at 500eV, half are at 1500eV and we use a Coulomb logarithm $\log \Lambda=1$. In Figure \ref{fig:FP}, we plot the distribution at the initial time and at two later times for both our solution using 20 polynomials and the Fokker-Planck result. The two solutions are completely indistinguishable from one another, indicating that, at least as far as the coefficients $S^n_{lk}$ and the Landau equation go, our computations are correct.
\subsection{Two-temperature plasma}
\begin{figure}
\begin{center}
\includegraphics[width=3.25in]{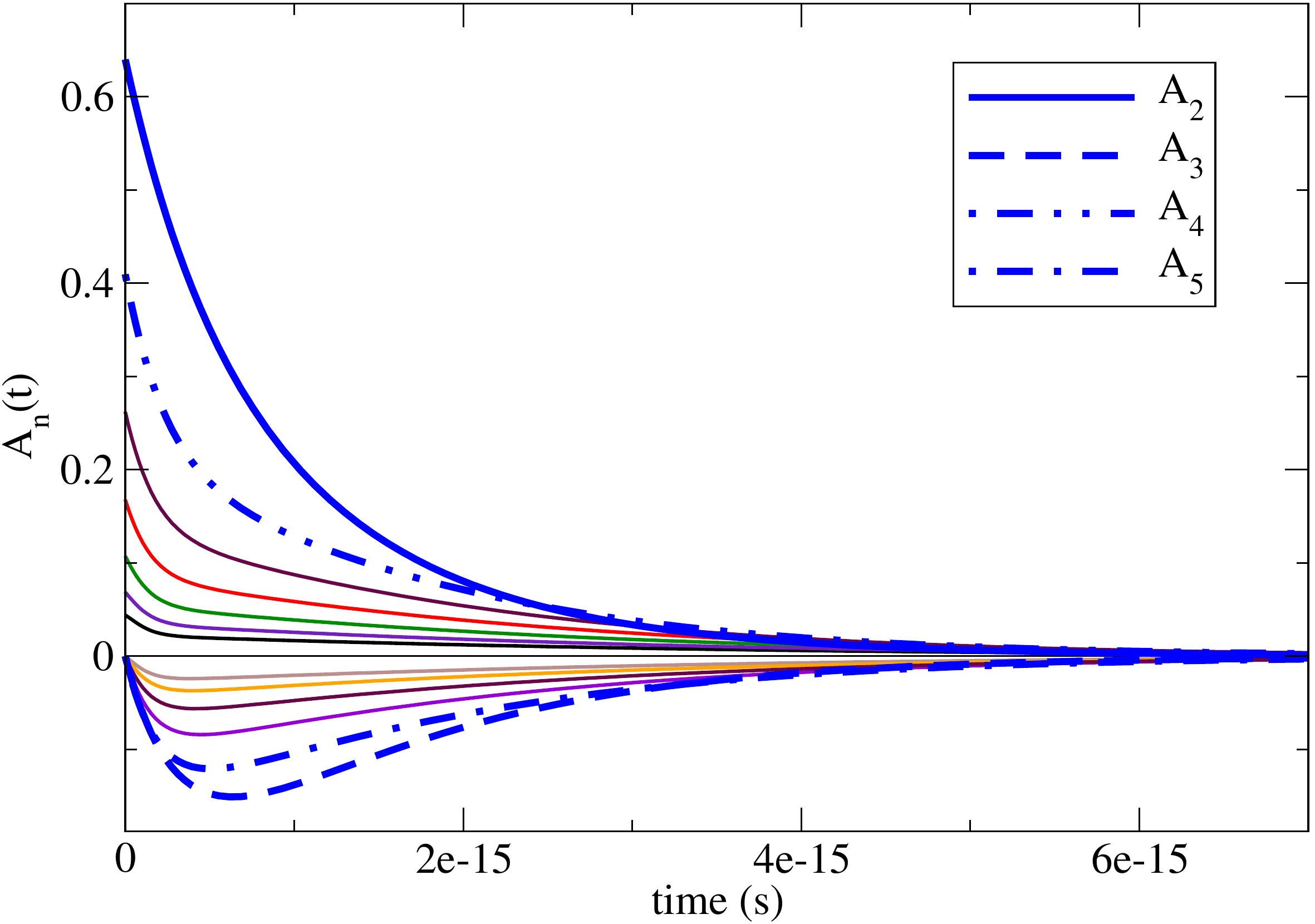}
\caption{Solution of the quantum Lenard-Balescu equation for a two-temperature initial condition with $\gamma=0.2$ and $T=1000$eV. The even coefficients, which all start out positive, decay monotonically to zero with the exception of $A_0$ which is fixed at 1. The odd coefficients start at zero and, aside from $A_1$, become negative and then decay to zero. The first four non-trivial coefficients are labelled and the rest up to $n=14$ are shown in various colors. The distribution itself is shown in Figure \ref{fig:distributionQLB}}.
\label{fig:gamma0.2_1000eV_QLB}
\end{center}
\end{figure}

\begin{figure}
\begin{center}
\includegraphics[width=3.25in]{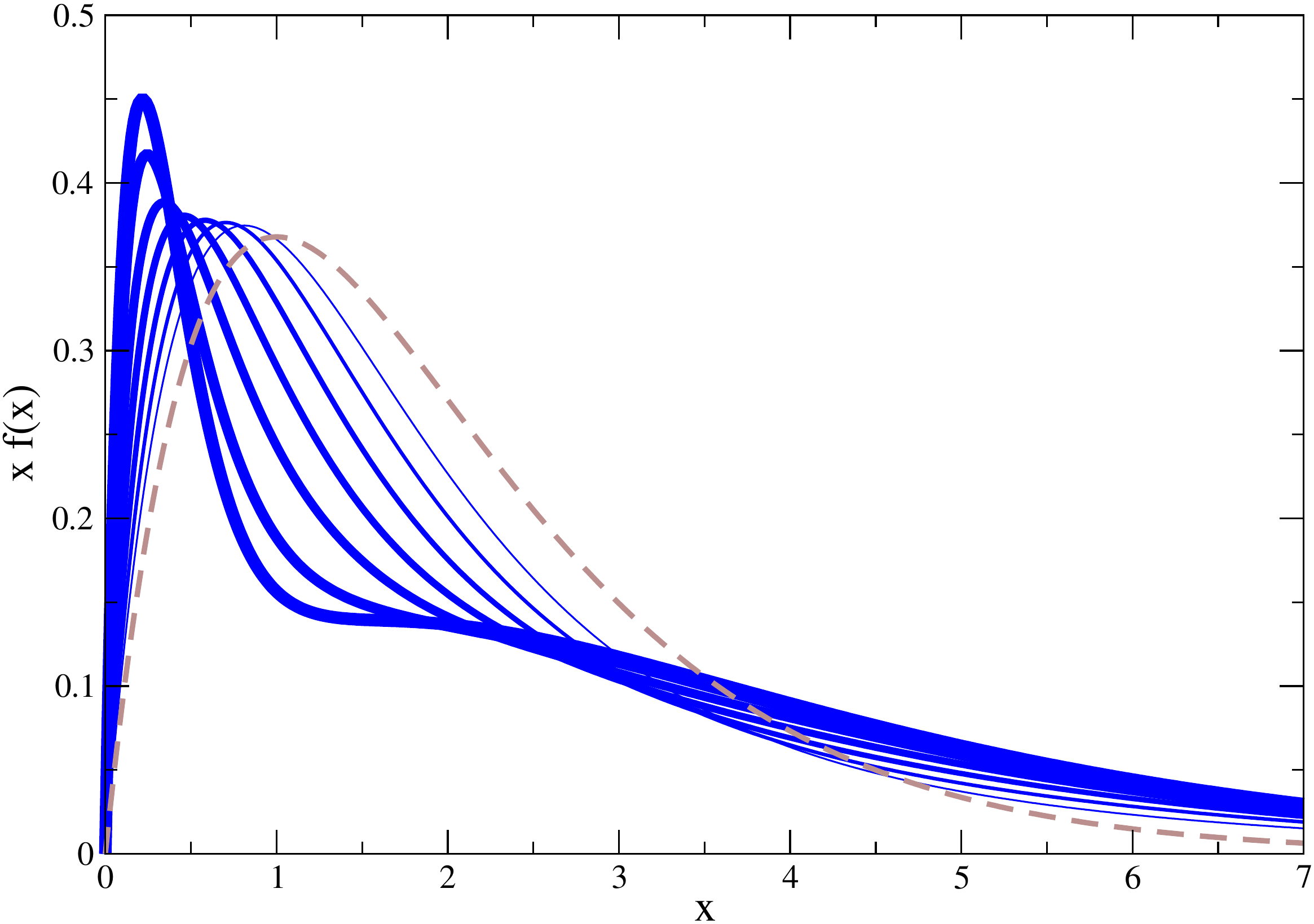}
\caption{Evolution of the distribution function under the quantum Lenard-Balescu equation for a two-temperature initial condition with $\gamma=0.2$ and $T=1000$eV. The thickest solid line is $t=0$ and the dashed line is equilibrium. The variable $x=\beta m_e v^2/2$ and $f(x)$ is divided by $n_e (\beta m_e/2 \pi)^{3/2}$.}
\label{fig:distributionQLB}
\end{center}
\end{figure}

\begin{figure}
\begin{center}
\includegraphics[width=3.25in]{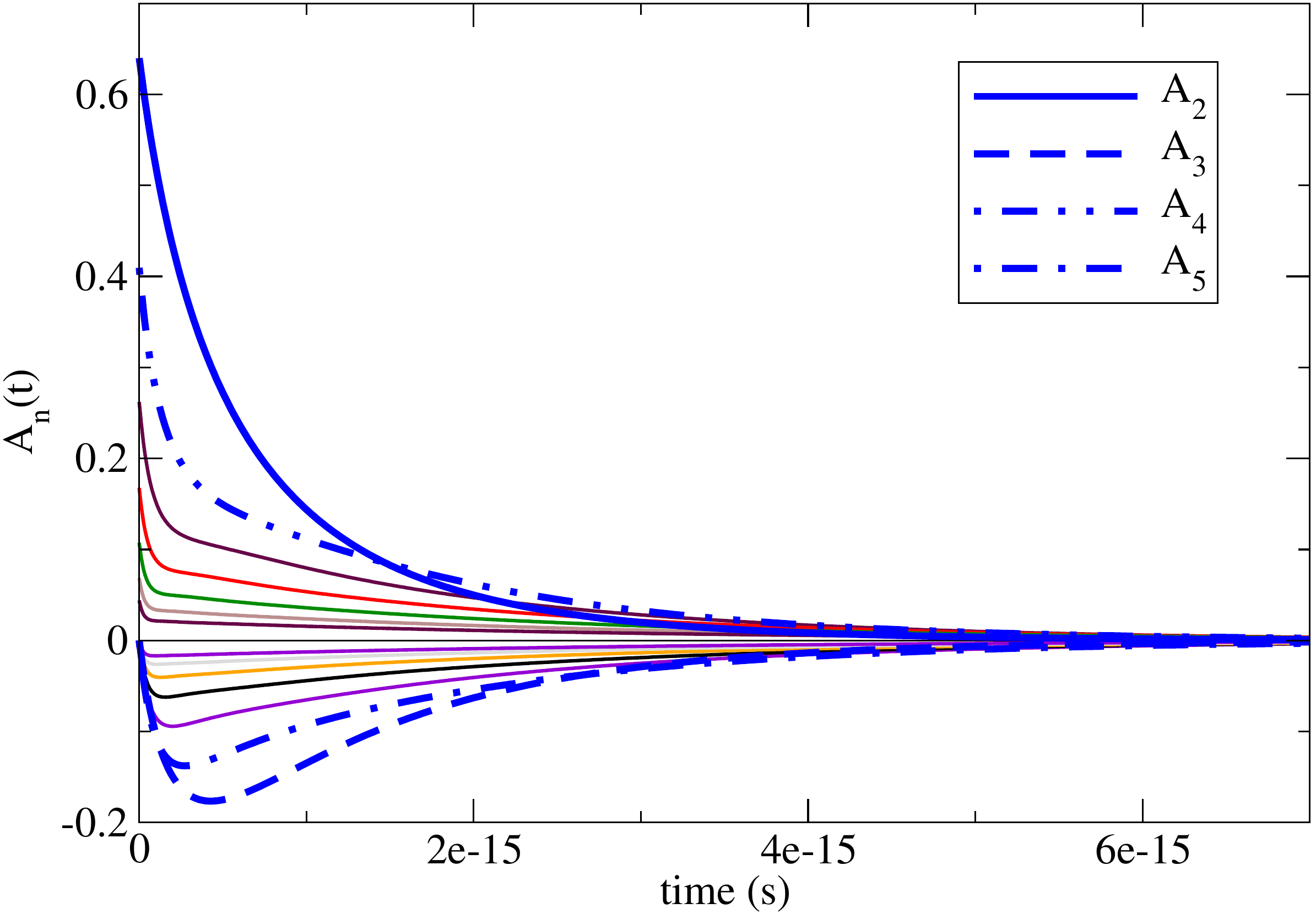}
\caption{Solution of the Landau equation for $\gamma=0.2$ and $T=600$eV.}
\label{fig:gamma0.2_600eV_Landau}
\end{center}
\end{figure}

\begin{figure}
\begin{center}
\includegraphics[width=3.25in]{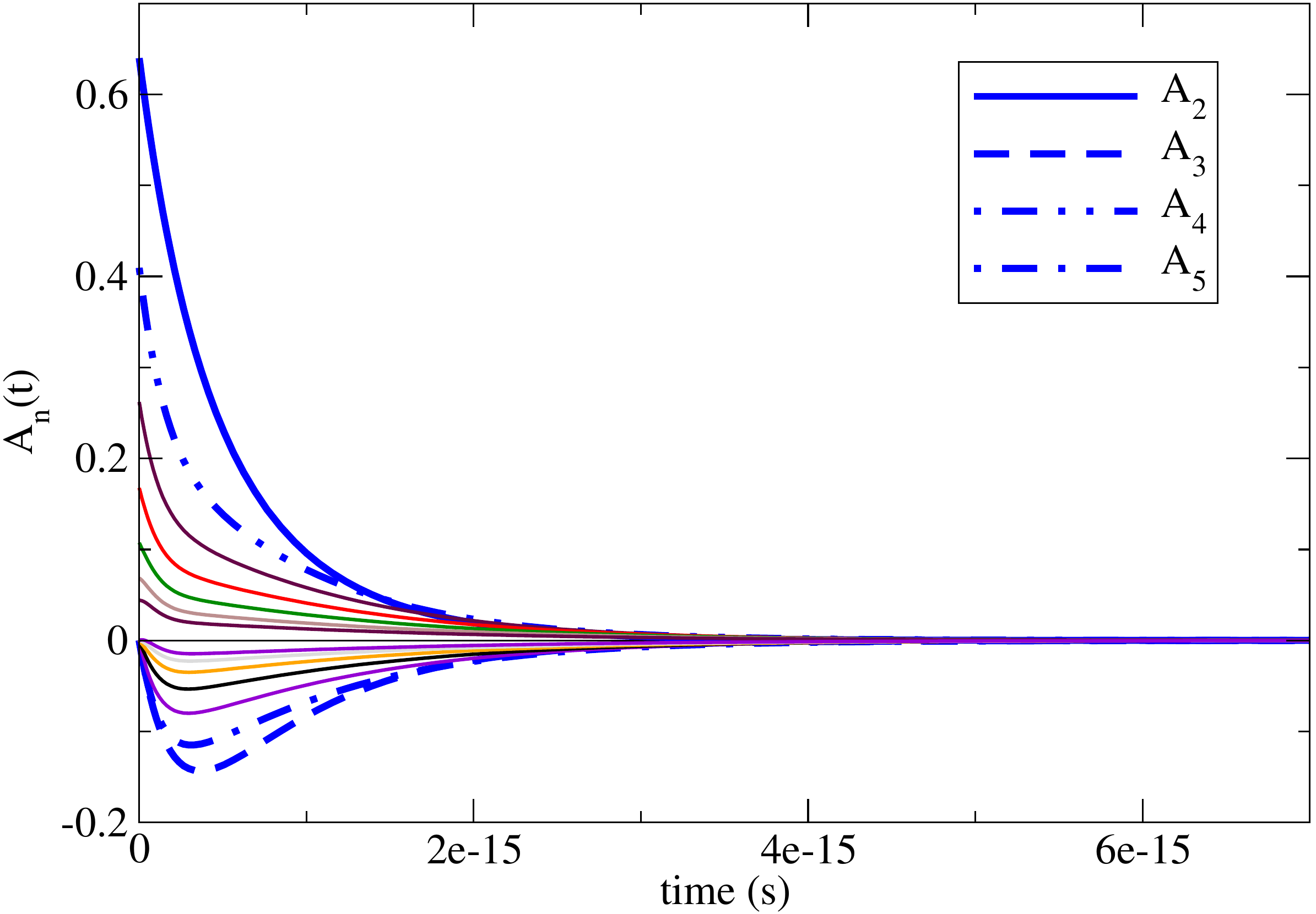}
\caption{Solution of the quantum Lenard-Balescu equation for $\gamma=0.2$ and $T=600$eV.}
\label{fig:gamma0.2_600eV_QLB}
\end{center}
\end{figure}

\begin{figure}
\begin{center}
\includegraphics[width=3.25in]{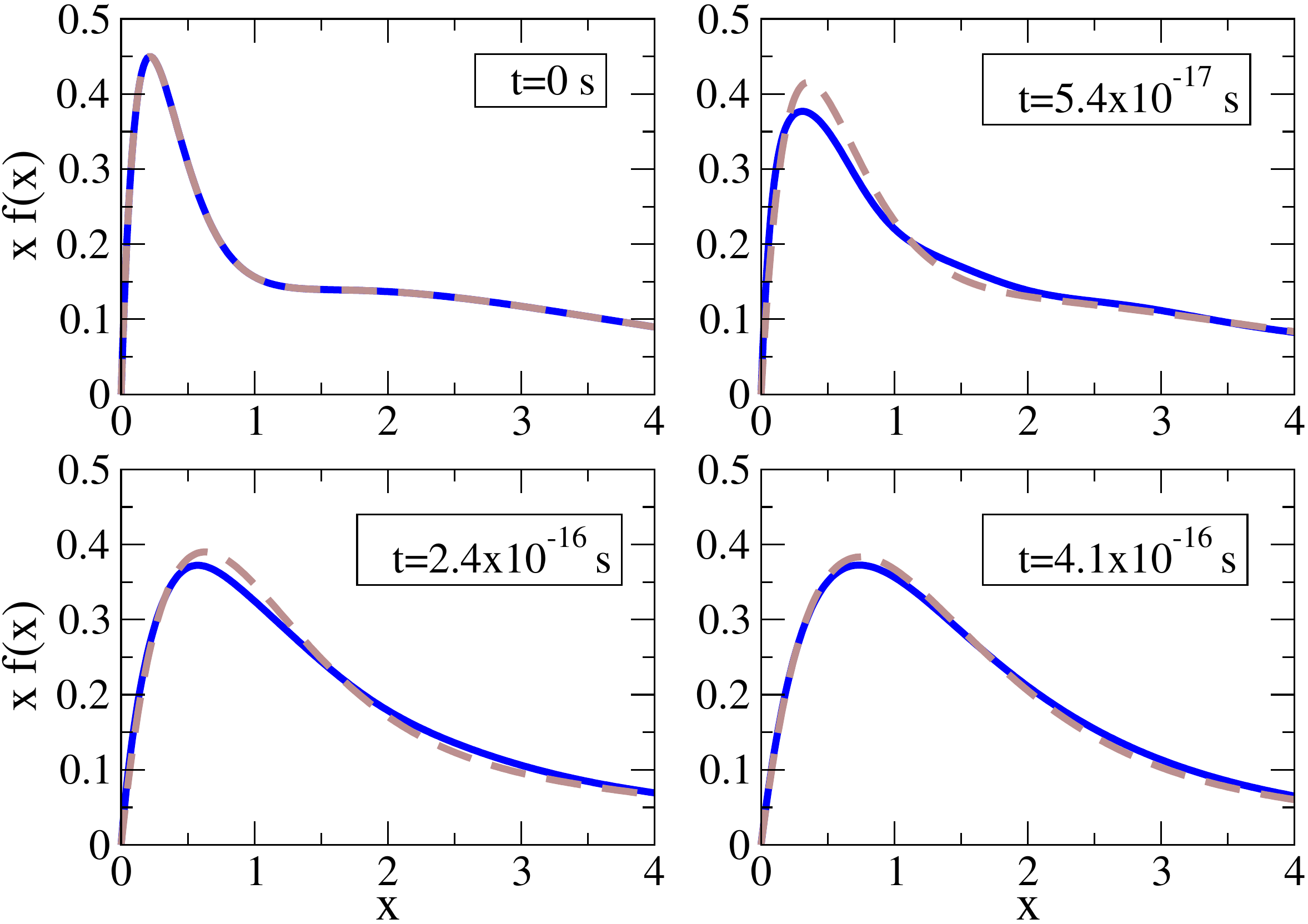}
\caption{Comparison between the evolution of the quantum Lenard-Balescu (solid blue) and Landau (dashed brown) equations for a two-temperature initial condition with $\gamma=0.2$ and $T=600$eV. The difference between the two is minimal at these conditions. The variable $x=\beta m_e v^2/2$ and $f(x)$ is divided by $n_e (\beta m_e/2 \pi)^{3/2}$.}
\label{fig:Landau_QLB_comp}
\end{center}
\end{figure}

Shown in Figure \ref{fig:gamma0.2_1000eV_QLB} is the numerical solution of the quantum Lenard-Balescu equation for the two-temperature plasma with $\gamma=0.2$, $n_e=1 \times 10^{25} \mathrm{cm}^{-3}$ and $T=1000$ eV. All the coefficients except for $A_0$, which is fixed at $1$, approach zero as $t \rightarrow \infty$, exactly as expected. The even coefficients fall monotonically while the odd coefficients, which start at zero, all become negative (except $A_1$ of course) before reaching a minimum and decaying back to zero. The distribution itself is shown in Figure \ref{fig:distributionQLB}. At 1000 eV, the solutions of the Landau and quantum Landau equations are essentially the same as Figures \ref{fig:gamma0.2_1000eV_QLB} and \ref{fig:distributionQLB}, indicating that the order-unity terms are not playing much role. This is not completely obvious since $\ln \eta$ is only around $-3$. As we reduce the magnitude of $\ln \eta$, which we do by turning down $T$, we can begin to see slight differences between the Landau and Lenard-Balescu solutions, although almost no difference is ever in evidence between the the Landau and quantum Landau equations. Shown in Figures \ref{fig:gamma0.2_600eV_Landau} and \ref{fig:gamma0.2_600eV_QLB} are the solutions of the Landau and quantum Lenard-Balescu equations for $T=600$eV, so $\eta \approx -2.3$; the evolution of the coefficients is noticeably different in the two cases. However, a comparison for the distribution itself is shown in Figure \ref{fig:Landau_QLB_comp}, and the differences between the Landau and quantum Lenard Balescu equations are modest at these conditions to say the least. We cannot turn the temperature down much further without having numerical problems in the solver, an indication that our neglect of higher-order terms in $\eta^2$ is becoming problematic. However, even at $T=600$eV, the low-temperature electrons are at 120eV and $5 \times 10^{24} \mathrm{cm}^{-3}$ and are becoming degenerate ($\theta \approx 1.1$). Thus, for this particular type of initial condition, our physical assumptions break down before we see any real advantage to carrying out the expensive integration over the dielectric function. On the other hand, we stress that any conclusions about where the Lenard-Balescu and Landau solutions become different are highly dependent on the initial distribution and we should not overestimate the generality of this particular example. It is certainly the case that by separating the temperature more widely, which we cannot do with only 40 polynomials, we would find ever greater divergence in the two solutions. In the next section, we find an initial distribution for which the two solutions are different.

\begin{figure}
\includegraphics[width=3.25in]{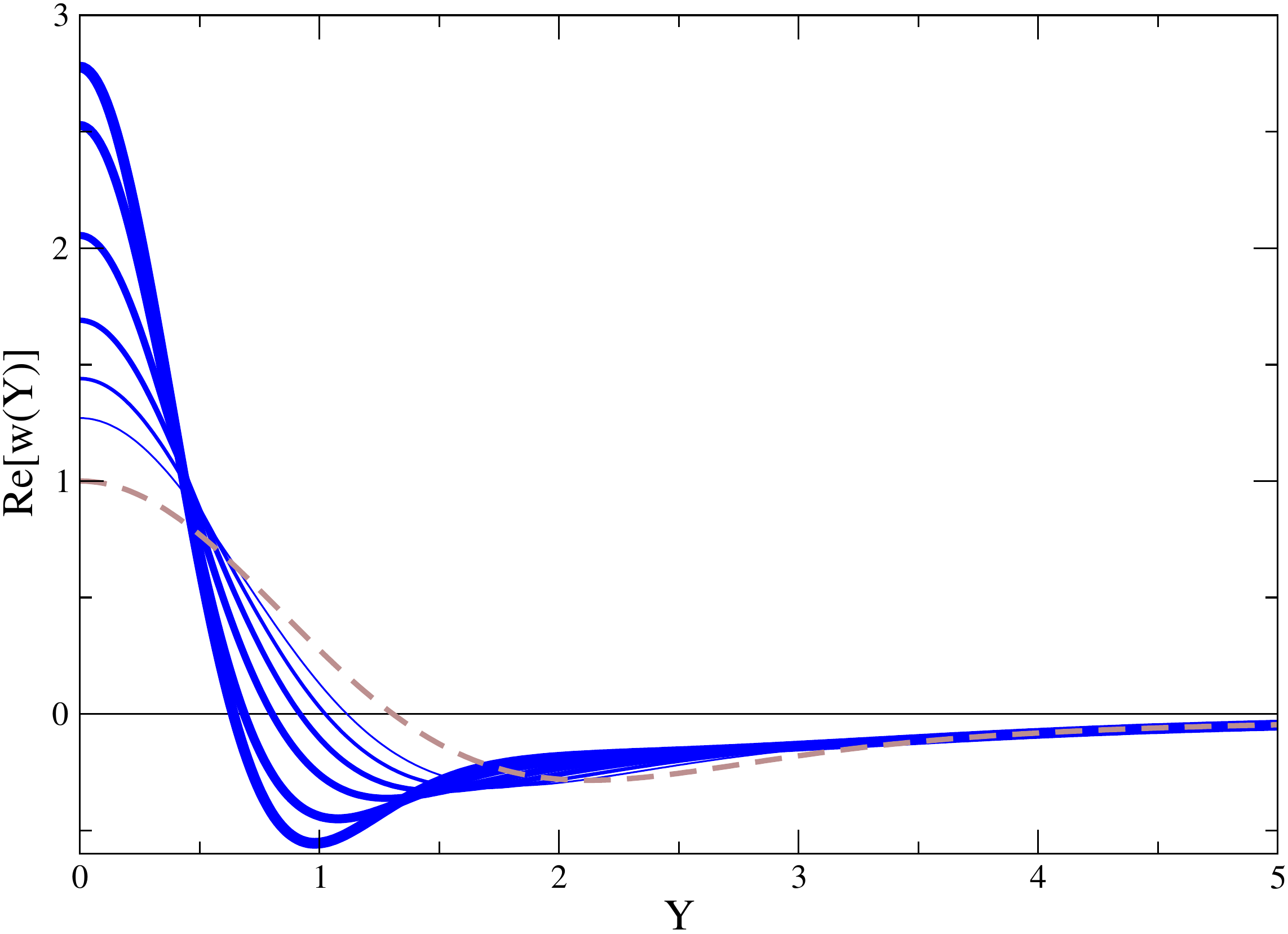}
\caption{Evolution of the real part of the free-particle response function under the quantum Lenard-Balescu equation for a two-temperature initial condition with $\gamma=0.2$ and $T=1000$eV. The thickest solid line is $t=0$ and the dashed line is equilibrium. The relationship to the response function in $k$ and $\omega$ space is $\chi(k,\omega)=-n_e \beta w(Y)$ where $Y$ is given by (\ref{eq:Y}).}
\label{fig:Rechi}
\end{figure}

\begin{figure}
\includegraphics[width=3.25in]{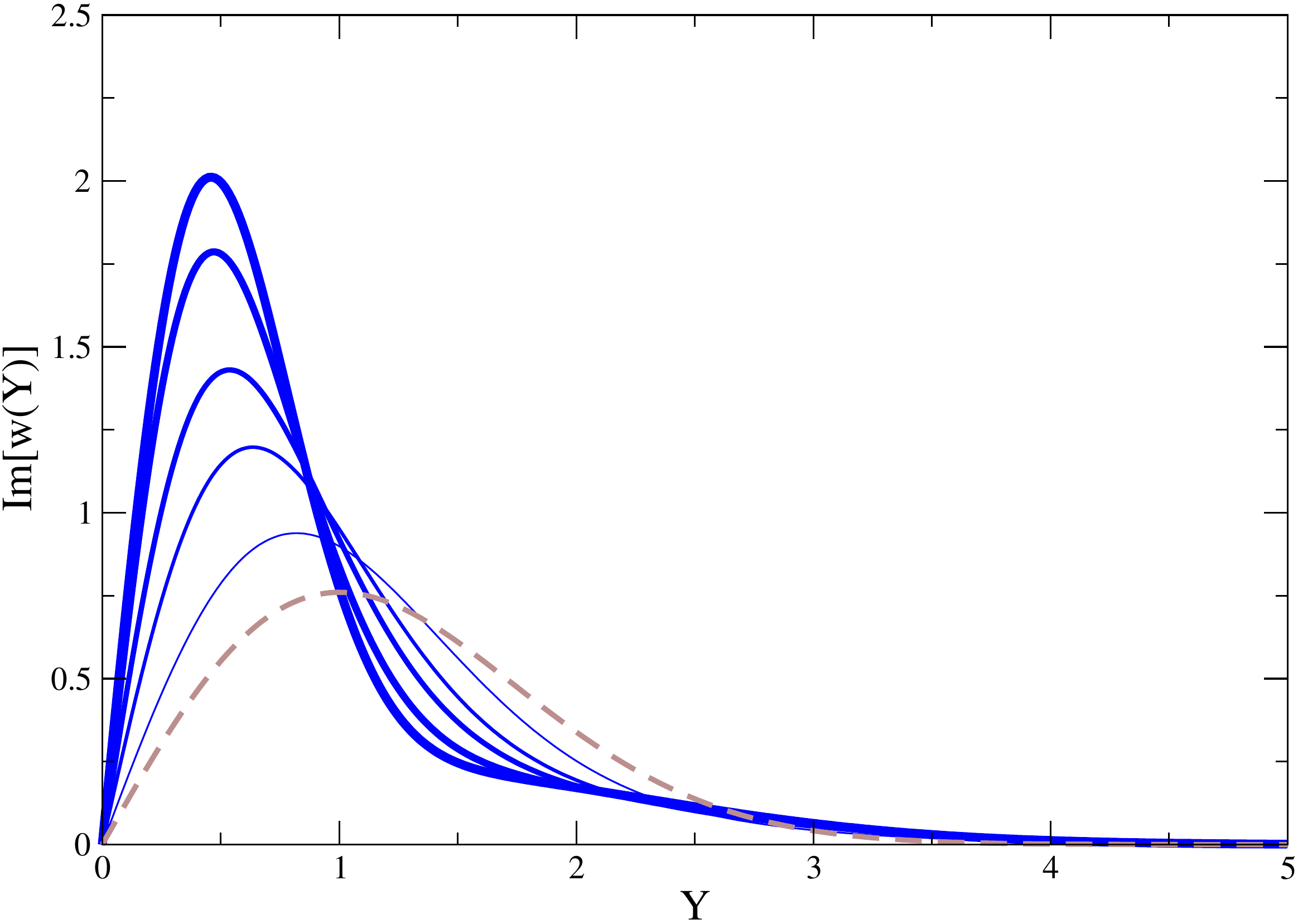}
\caption{Evolution of the imaginary part of the free-particle response function under the quantum Lenard-Balescu equation for a two-temperature initial condition with $\gamma=0.2$ and $T=1000$eV. The thickest solid line is $t=0$ and the dashed line is equilibrium. The relationship to the response function in $k$ and $\omega$ space is $\chi(k,\omega)=-n_e \beta w(Y)$ where $Y$ is given by (\ref{eq:Y}).}
\label{fig:Imchi}
\end{figure}

Our method also allows a detailed view of the dielectric function in the random phase approximation, something that would not be easy to obtain with a discretization method. Figures \ref{fig:Rechi} and \ref{fig:Imchi} show the time evolution of the real and imaginary parts of the free-particle response function for the two-temperature initial condition with $\gamma=0.2$, $T=1000$eV and $n_e=1 \times 10^{25} \mathrm{cm}^{-3}$. These are easily obtained from the coefficients and equations (\ref{eq:wr}) and (\ref{eq:wi}).
\subsection{Under-resolved two-temperature plasma}
\begin{figure}
\begin{center}
\includegraphics[width=3.4in]{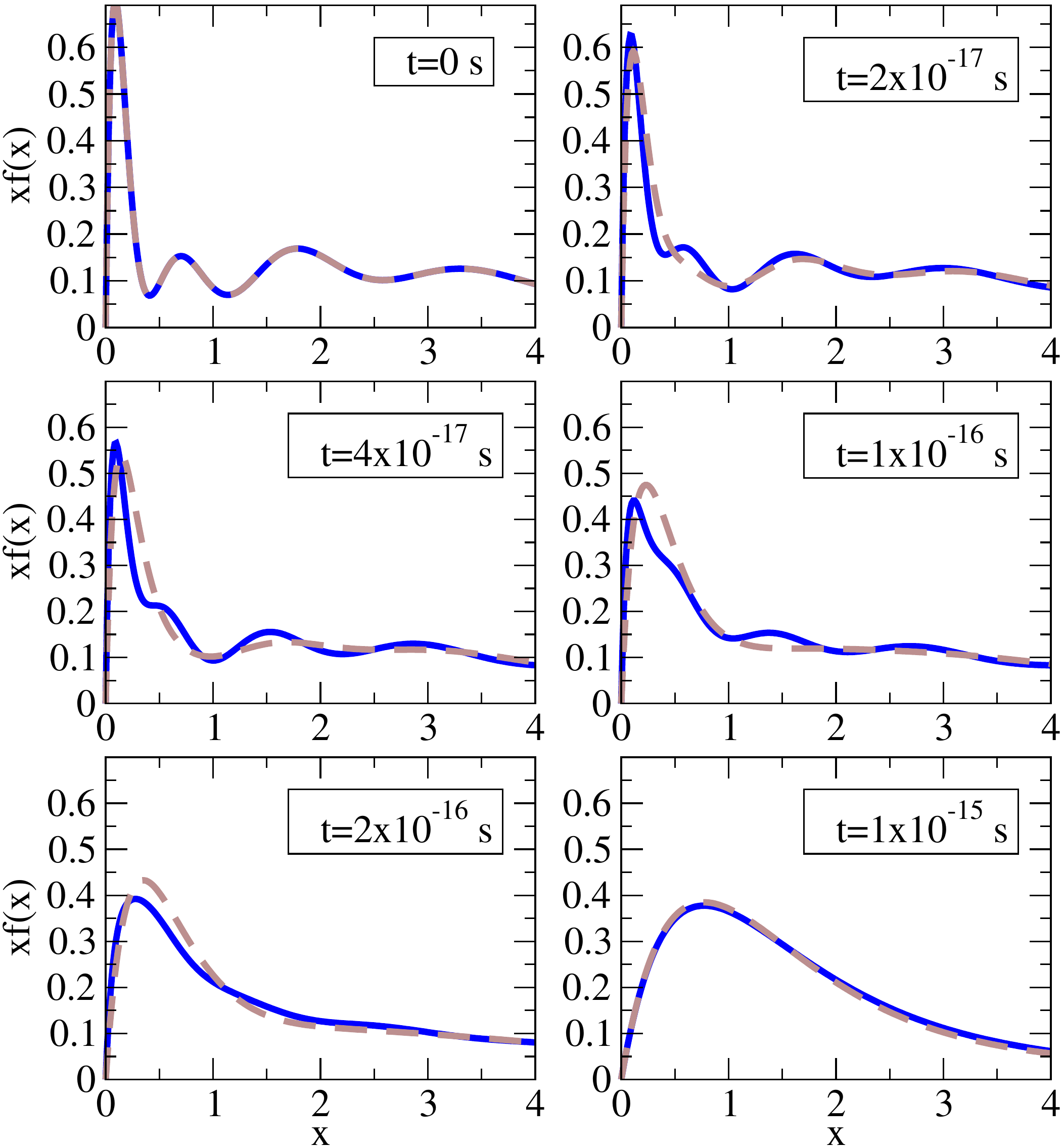}
\caption{Comparison between the quantum Lenard-Balescu (solid) and Landau (dashed) equations for a two-temperature initial condition with $\gamma=0.08$ and $T=1000$eV. We do not have enough polynomials to fully resolve this distribution but the series expansion still constitutes a valid initial condition. The variable $x=\beta m_e v^2/2$ and $f(x)$ is divided by $n_e (\beta m_e/2 \pi)^{3/2}$.}
\label{fig:gamma0_08comp1000eV}
\end{center}
\end{figure}

\begin{figure}
\begin{center}
\includegraphics[width=3.4in]{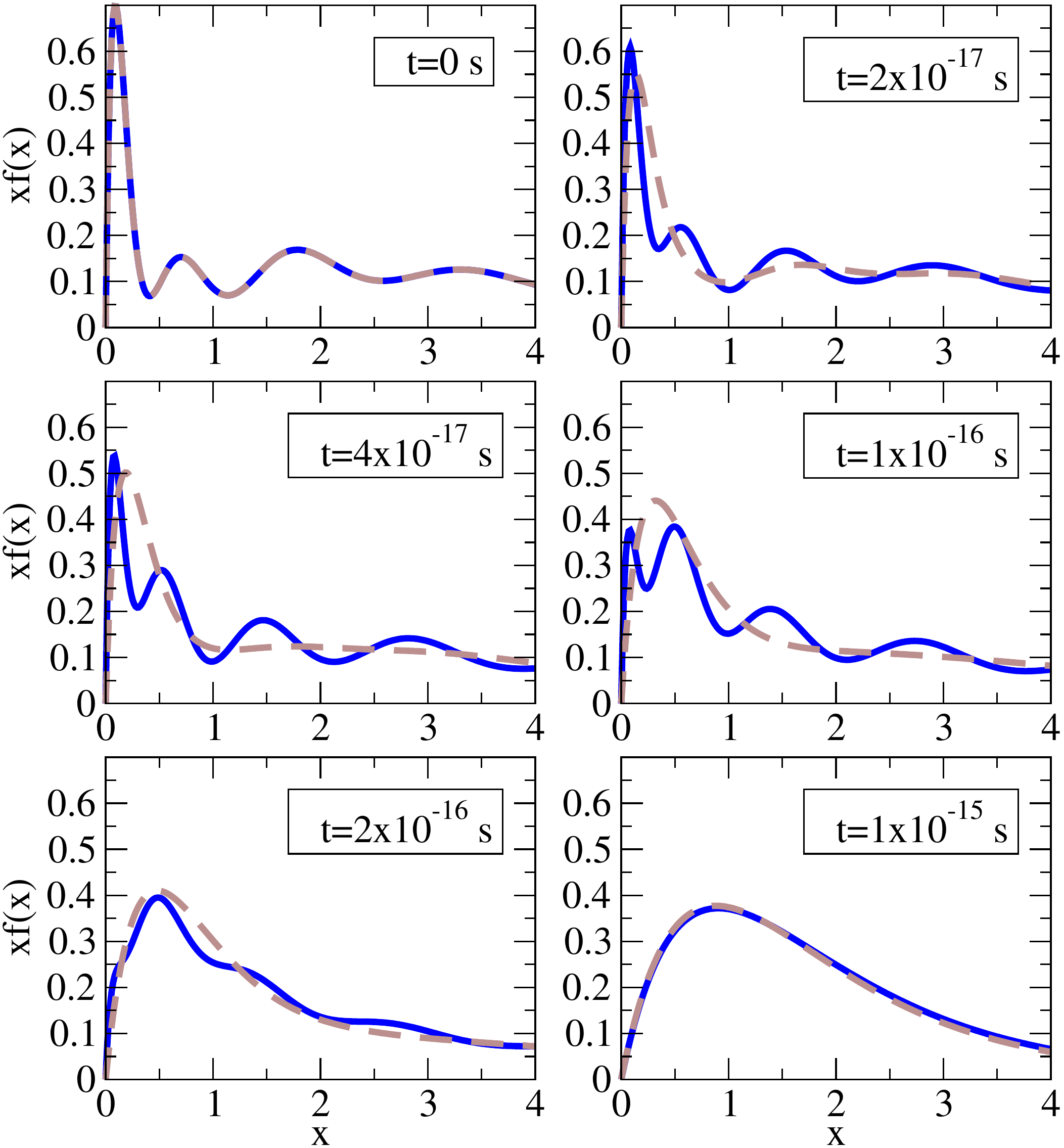}
\caption{Comparison between the quantum Lenard-Balescu (solid) and Landau (dashed) equations for a two-temperature initial condition with $\gamma=0.08$ and $T=600$eV. We do not have enough polynomials to fully resolve this distribution but the series expansion still constitutes a valid initial condition. The variable $x=\beta m_e v^2/2$ and $f(x)$ is divided by $n_e (\beta m_e/2 \pi)^{3/2}$.}
\label{fig:gamma0_08comp}
\end{center}
\end{figure}

Here, we use the two-temperature initial condition but choose $\gamma=0.08$, which is much too small for 40 polynomials. However, as Figure \ref{fig:twoTinit} shows, even though this distribution is badly under-resolved, it is still positive everywhere and thus constitutes a viable initial condition. To solve this problem, we keep the first $34$ polynomials for the initial condition and set the remaining seven to zero. This way we still maintain a positive distribution but we have a few modes above our resolved range to ensure we have sufficient resolution for the subsequent evolution. We find that for $T=1000$eV, there are modest but clear differences between the Landau and QLB evolutions, as shown in Figure \ref{fig:gamma0_08comp1000eV}. At $600$eV, the two solutions are very different, as shown in Figure \ref{fig:gamma0_08comp}. The Landau equation more quickly smoothes out the ripples in the distribution than QLB and we have two very distinct approaches to equilibrium.


\subsection{Gaussian distribution}
\begin{figure}
\begin{center}
\includegraphics[width=3.25in]{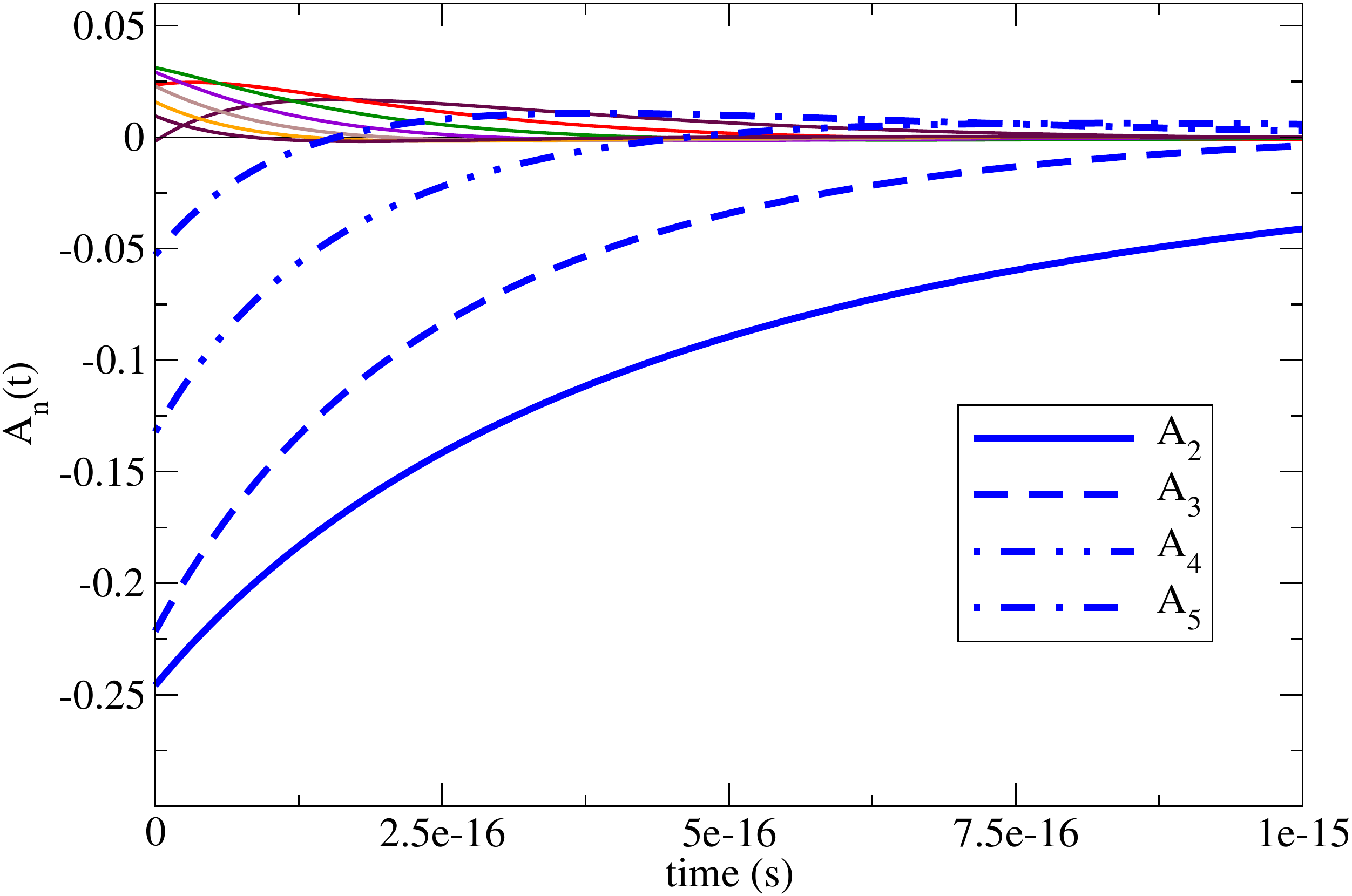}
\caption{Evolution of the first few $A_n$ for the Gaussian initial distribution, equation (\ref{eq:gaussian}).}
\label{fig:gaussian}
\end{center}
\end{figure}

\begin{figure}
\begin{center}
\includegraphics[width=3.25in]{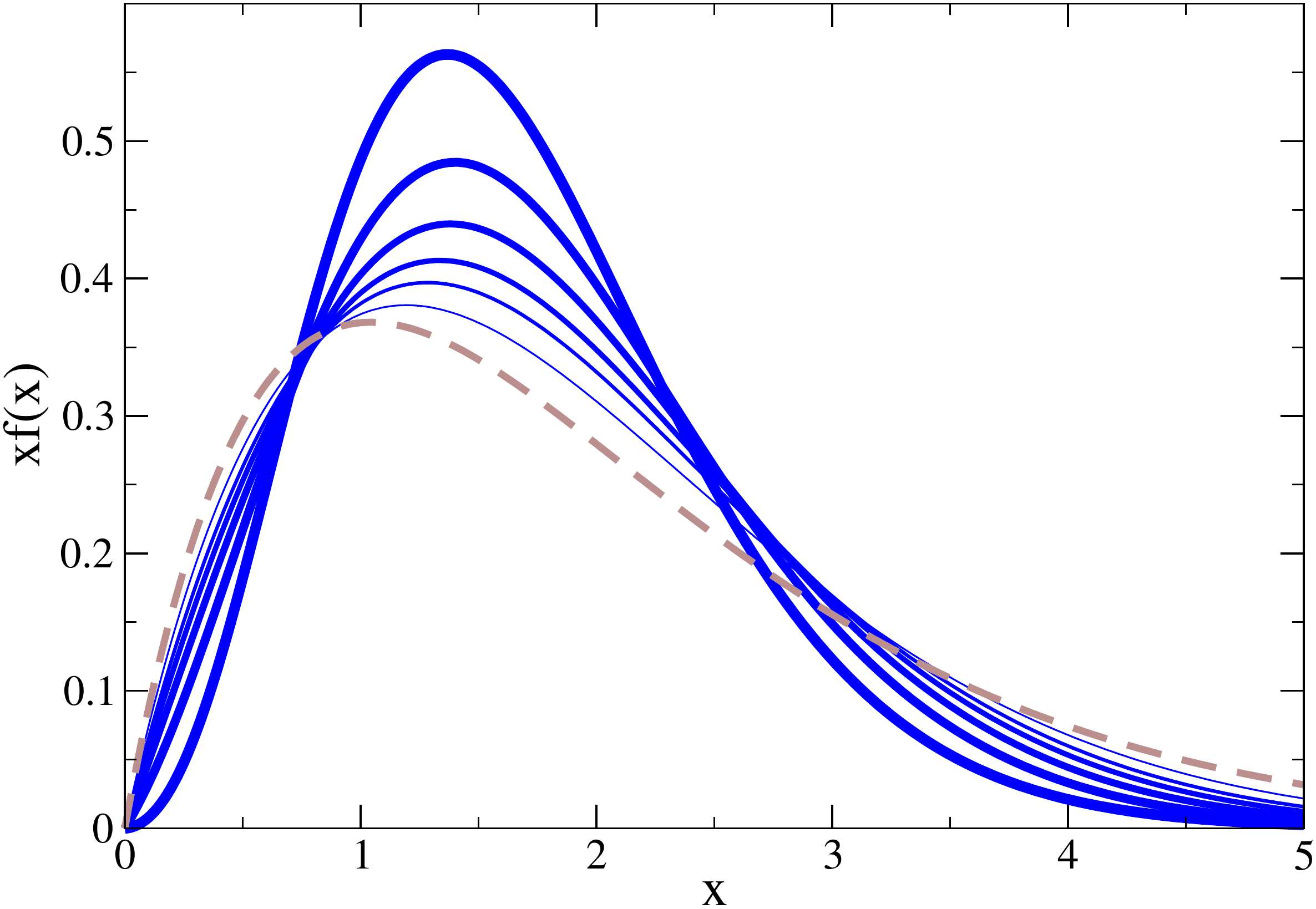}
\caption{Evolution of Gaussian initial distribution, equation (\ref{eq:gaussian}) with $\bar{v}_0=2$ and $T=1000$eV. The thickest line is $t=0$ and the dashed line is equilibrium. The variable $x=\beta m_e v^2/2$ and $f(x)$ is divided by $n_e (\beta m_e/2 \pi)^{3/2}$.}
\label{fig:gaussian_distribution}
\end{center}
\end{figure}

We solve for the relaxation of the Gaussian initial condition described in Section \ref{sec:Gauss_init} with $T=1000$eV, $v_0=2$ and $n_e=1.0 \times 10^{25} \mathrm{cm}^{-3}$. The evolution of the coefficients is shown in Figure \ref{fig:gaussian}, while that of the distribution itself is in Figure \ref{fig:gaussian_distribution}. As in the two-temperature case, there is not much difference between the Landau and quantum Lenard-Balescu equations at these conditions. And once again, upon making $\eta$ smaller, our physical and numerical approximations break down before we see any interesting differences.

\section{Generalizations and variations} \label{sec:generalizations}

\subsection{Beyond order unity}
All the calculations we have done here have the logarithmic and order unity terms. To get all the higher order terms is a straightforward generalization. Consider the integral in equation (\ref{eq:Cnlk}),
\begin{equation}
T^n_{lk} \equiv \int_0^{\infty} dX \frac{e^{-X^2}}{X^3} \int_{-\infty}^{\infty} dY \frac{e^{-Y^2}}{|\epsilon(X,Y)|^2} P^n_{lk}(X,Y). \label{eq:Tnlk}
\end{equation}
We can decompose the polynomial into powers of $X$,
\begin{equation}
 P^n_{lk}(X,Y)= \sum_{p=1}^M X^{2p} G^{np}_{lk}(Y) \label{eq:Pdecomp}
\end{equation}
and keep the dielectric function everywhere, using the formulas in Appendix C (with $p=s+1$) to do the $X$-integrals for every $p$ rather than just $p=1$ as we have done. Using the decomposition (\ref{eq:Pdecomp}) and the results of Appendix C, the integral (\ref{eq:Tnlk}) can be written
\begin{equation}
 T^n_{lk}=-\frac{1}{2} \sum_{p=0}^M p! \eta^{2p-2} J_p(\eta) \label{eq:Tnlkexact}
\end{equation}
where
\begin{eqnarray}
 i J_p(\eta)&\equiv& \int_{-\infty}^\infty \frac{[w(Y)]^p}{w_i(Y)} e^{-Y^2} e^{\eta^2 w(Y)} G^{np}_{lk}(Y) \times \cr 
& &\ \ \ \ \ \ \ \Gamma\left(-p,\eta^2 w(Y) \right) dY . \label{eq:Jp}
\end{eqnarray}
Although not clear by inspection, the real part of the integrand in (\ref{eq:Jp}) is odd for all $p$ so the integral is always imaginary. Calculating these $Y$-integrals numerically by the method we used in this work would require that we precompute $G^{np}_{lk}(Y)$ at the quadrature points for all $(n,l,k)$ and all $p$ up to $M$, which varies depending on the polynomial. This would lead to a large quantity of precomputed data but it is possible in principle. Alternatively, we can keep the dielectric function for $1 \le p \le p_\mathrm{max}$ but set it to $1$ for $p>p_\mathrm{max}$, which would allow us to keep a prescribed number of powers of $\eta$. If the method of Appendix E, or something like it, proves feasible then we could use it to compute all the coefficients after putting our effort into computing the coefficients for $l=0$ alone. This would probably be the ideal solution if it is possible.

\begin{figure}
\begin{center}
\includegraphics[width=3.25in]{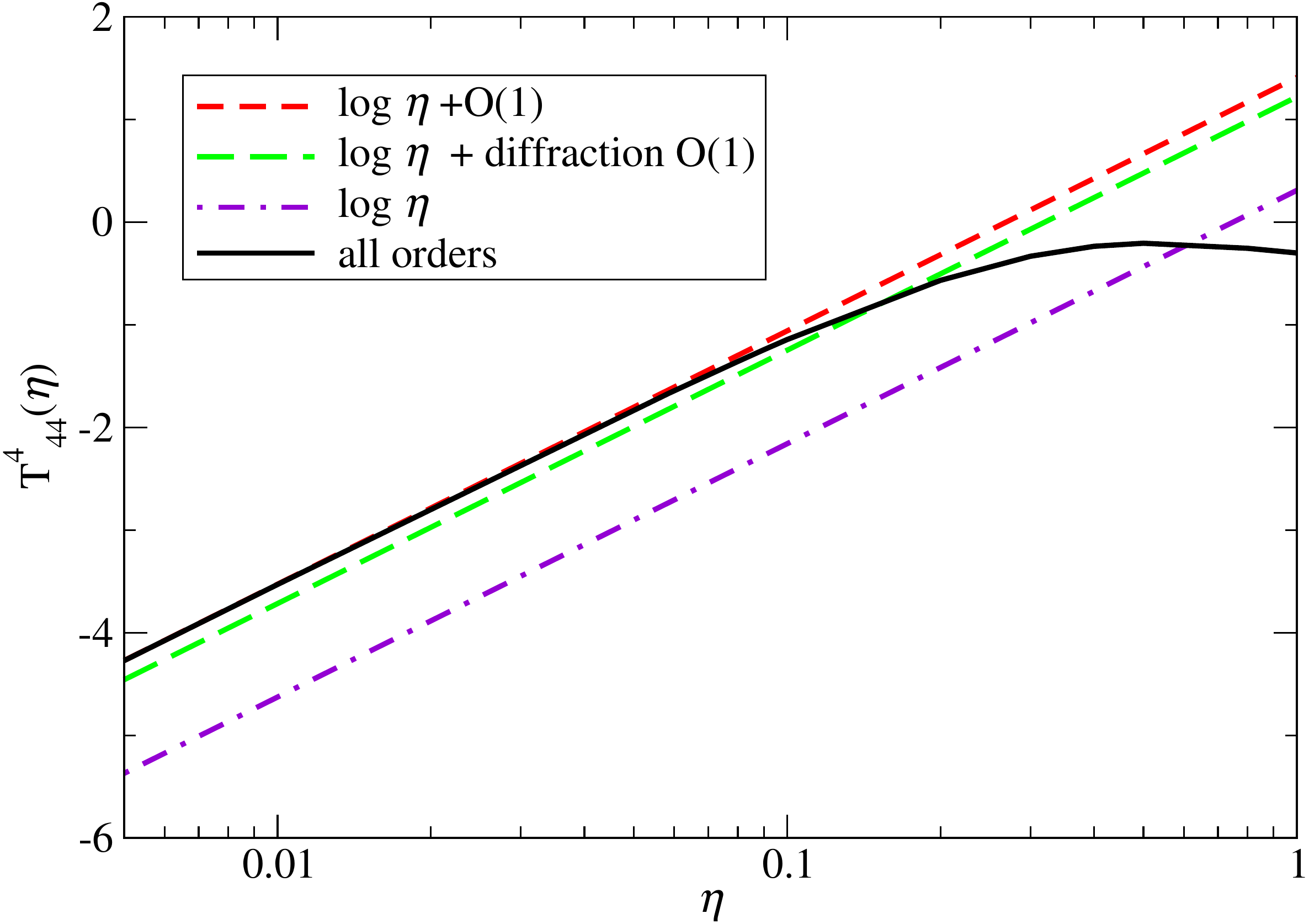}
\caption{The integral $T^4_{44}(\eta)$ defined in equation (\ref{eq:Tnlk}). The solid black curve is the integration carried out to all orders in $\eta$; short dashed red is the approximation  (\ref{eq:Cfinal}) used in our solution of the quantum Lenard-Balescu equation; long dashed green uses (\ref{eq:Cfinal}) with $F^n_{lk}=0$; dash-dotted purple uses (\ref{eq:Cfinal}) with neither $B^n_{lk}$ or $F^n_{lk}$. Our approximation is very accurate until $\eta \sim 0.1$ and is much better than using the Coulomb logarithm alone.}
\label{fig:T444}
\end{center}

\end{figure}

To examine where higher powers of $\eta$ might be needed, we compute $T^n_{lk}(\eta)$ for $(n,l,k)=(4,4,4)$ using the formula (\ref{eq:Tnlkexact}). We compute the $Y$-integrals with Mathematica's adaptive numerical integration for the case of a two-temperature plasma with $\gamma=0.5$ and $20$ polynomials. This calculation discards no powers of $\eta$ and we compare it with the approximation used in the solution of the QLB equation,
\begin{equation}
 T^n_{lk} \approx -\frac{\Gamma(n+3/2)}{n!}\left[S^n_{lk} \left(\frac{\gamma_E}{2}+\ln \eta \right) - B^n_{lk} + F^n_{lk} \right] \ . \label{eq:Tapprox}
\end{equation}
The result is shown in the short-dashed red curve in Figure \ref{fig:T444}. Our approximation is very accurate until $\eta \sim 0.1$ and then higher-order terms become necessary. We stress again that this conclusion is highly dependent upon the distribution, but among the ones we are able to resolve with $40$ polynomials, this gives a reasonable idea of where the approximations start to break down. We also compare with the result of setting either $B^n_{lk}$, $F^n_{lk}$ or both, to zero. The dash-dotted purple line indicates that the Landau approximation, where both these terms are set to zero, is worst. Keeping $B^n_{lk}$ but not $F^n_{lk}$ gives the long-dashed green curve, which is a marked improvement, but keeping both is clearly best and is very accurate when $\eta<0.1$. Table \ref{table:T444} gives the actual values for $\eta=0.001$. Now, for this distribution at these conditions there is not much difference between the solutions of the Landau and QLB equations, so the discrepancies in Figure \ref{fig:T444} apparently do not have a noticeable effect.

\begin{table}
\begin{center}
 \begin{tabular}{|c|c|}
 \hline 
 Approximation & $T^4_{44}(0.001)$ \\
 \hline
 All orders & -5.99681 \\
 Equation (\ref{eq:Tapprox}) & -5.99678 \\
 Quantum Landau & -6.1849 \\
 Landau & -7.09642 \\
 \hline
 \end{tabular}
\caption{Comparison of the various approximations for calculating $T^4_{44}(\eta)$ defined in equation (\ref{eq:Tnlk}). Quantum Landau is equation (\ref{eq:Tapprox}) with $F^n_{lk}=0$ and Landau is (\ref{eq:Tapprox}) with $B^n_{lk}=F^n_{lk}=0$.}
 \label{table:T444}
\end{center}
\end{table}

\subsection{Multiple species}
The generalization to multiple species is straightforward and the techniques for the evaluation of the dielectric function integrals will work in that case too. If we have, say, electrons and protons, we would need two sets of coefficients $A_n^e$ and $A_n^p$ corresponding to the expansion (\ref{eq:expansion}) for $f_e(v,t)$ and $f_p(v,t)$. Of course, there would also be collision operators for $e-e$, $p-p$ and $e-p$ interactions. If we choose $u=1$, conservation of particles is given by $A^e_0=A^p_0=1$ and energy by $A_1^e=-A_1^p$.

\subsection{Velocity anisotropy and spatial inhomogeneity}
To treat the most general Wigner distributions, $f({\bf r},\bfv,t)$, we can likewise generalize the expansion (\ref{eq:expansion}) to be
\begin{equation}
f^{\mathrm{eq}}({\bf r},{\bf v},t) \sum_{nlm} A_{nlm}({\bf r},t) Y_{lm}(\theta,\phi) \Lnhalf \left(\frac{u \beta m_e v^2}{2} \right),
\end{equation}
where $Y_{lm}(\theta,\phi)$ are spherical harmonics. The $A_{nlm}({\bf r},t)$ now satisfy partial differential equations and conservation laws are given by integrals of the coefficients over space, which must be respected by the solution method. The results of Appendices A and B must also be generalized to include the spherical harmonics. This does not appear to be particularly easy, and we may find ourselves lacking the convenient closed formulas we obtained in the isotropic case, but it is surely not impossible.

\subsection{Alternative expansions}
There are other related expansions that can be used to solve these equations. First, we explore the possibility of choosing different values of $u$. As mentioned in section \ref{sec:method}, we have been content here to use $u=1$ because of the trivial conservation properties (\ref{eq:particles}) and (\ref{eq:energy}) but we cannot represent every possible distribution this way. This situation can be rectified by keeping $u$ in equation (\ref{eq:expansion}). The essential form of such an expansion is
\begin{equation}
 f(x)=e^{-x/u} \sum_{n=0}^\infty A_n \Lnhalf(x),
\end{equation}
that is, the argument of the exponential is a factor of $u$ smaller than that of the Laguerre polynomial. For this expansion to converge, $e^{x/u}f(x)$ must be square-integrable with respect to the Laguerre weight, so
\begin{equation}
 \int_0^\infty x^{1/2} e^{x(-1+2/u)} |f(x)|^2 dx < \infty \ .
\end{equation}
If $u=2$ this condition is very mild, certainly far less stringent than (\ref{eq:sqint}). The cost of this is that we complicate the collision integrals even further, and conservation is no longer automatic. In Appendices A and B, we indicate the modifications needed for computing the response function and collision integrals with general $u$. The condition for conservation of particles becomes 
\begin{equation}
 \frac{2}{\sqrt{\pi}}\sum_{n=0}^\infty A_n \frac{\Gamma(n+3/2)}{n!} (1-u)^n=1
\end{equation}
and for conservation of energy we have,
\begin{equation}
 \frac{2}{3 \sqrt{\pi}} \sum_{n=0}^\infty A_n \frac{\Gamma(n+3/2)}{n!} (3-3u-2nu) (1-u)^n=1 \ .
\end{equation}
Thus, if $u=2$, conservation is no longer automatic but can be lost if we have an insufficient number of coefficients. Of course, one may question whether it is particularly valuable to maintain conservation of particles and energy even if we have insufficient resolution, so this may not be a very important consideration. Also, even though $u=1$ was sufficient for our purposes because the system does not evolve to a state that violates (\ref{eq:sqint}) if it is initially satisfied, when we generalize this method to multiple particle species and situations in which there is an external force, this may no longer be the case. Therefore, in more practical applications, it may be that $u=2$, or at least $u>1$, is more appropriate.

An expansion based on a completely different set of orthogonal functions may prove useful. Rather than the orthogonality condition (\ref{eq:orth}), one might be tempted to try polynomials that satisfy
\begin{equation}
 \int_0^\infty x^\nu e^{-x^2} P_n(x) P_m(x) dx = M_n \delta_{nm} \label{eq:orthMaxwell}
\end{equation}
where $M_n$ are a set of constants. The distribution expansion is then
\begin{equation}
 f(v,t)=n_e \left(\frac{m_e \beta}{2 \pi}\right)^{\frac{3}{2}} e^{-\frac{\beta m_e v^2}{2}}\sum_{n=0}^{\infty} A_n(t) P_n \left(\sqrt{m_e \beta/2} v\right) \ .
\end{equation}
In recent work \cite{Landreman,Wilkening}, it was suggested that these polynomials may be a more efficient way to represent distribution functions for certain applications. It does seem to be the case \cite{Baddoo} that, compared with the Laguerre polynomials, one needs fewer of them to fully resolve some distributions (in our case, $\nu=2$ is the natural choice). However, these polynomials are ``non-classical'' \cite{Miranian} and there are no closed forms either for the polynomials themselves or for any of their properties such as coefficients of the recurrence relation and the normalization constants, $M_n$. The polynomials must be generated by the Gram-Schmidt procedure and all required quantities, such as the real and imaginary parts of the response function and the polynomial $G^n_{lk}(Y)$ would need to be computed numerically at the quadrature points without the aid of any of the exact formulas upon which we have relied. Of course, such closed forms are not really necessary and it may be worth exploring this issue further.

\section{Discussion}
We have used a spectral expansion to solve the quantum Lenard-Balescu equation for a one-component Coulomb system. To demonstrate the technique, we have computed the relaxation to equilibrium of various initial distributions including variants of a two-temperature plasma. We have found that including the full dynamical dielectric function makes little difference for these problems and we do just as well if we use the computationally cheaper static screening. This is in general agreement with the findings of Dolinsky for the classical Lenard-Balescu equation; he was not able to find an initial condition for which there was any difference between the Lenard-Balescu and Landau/Fokker-Planck systems. Nevertheless, this conclusion about the relevance of the dielectric function cannot be true in general. As we pointed out, even for a one-component plasma divided into two temperatures, with a large enough temperature ratio significant differences should be expected between the two kinetic solutions. For the moment, we do not have the resolution to thoroughly study this effect but we began to see hints of it in our under-resolved $\gamma=0.08$ solutions. When we generalize to multiple species and anisotropic distributions, the dielectric function may become more important. For example, one can make a significant error in calculations of thermal conductivity for an electron-proton system by using static over dynamical screening in the collision integrals. This difference will also be present in time-dependent solutions.

Compared to the many advances for the Boltzmann and Landau/Fokker-Planck equations \cite{Tzoufras,Taitano2015,Taitano2015-2,Haack2012}, for which it is now possible to find solutions in multiple dimensions of velocity and space, our velocity-isotropic and 0D spatial solutions may not seem terribly impressive. However, we have shown that it is possible to solve the quantum Lenard-Balescu equation including faithful integrations over the dielectric function. What is more, our method provides an analytic representation of the response and dielectric functions. It is also readily generalizable to multiple space and velocity dimensions and we hope that this will be the subject of future work.

\section{Acknowledgments}
We are grateful to David Michta for providing data for comparison with the Landau solution and to Lorin Benedict, Michael Murillo, Antoine Cerfon and Cory Hauck for useful discussions. Part of this research was performed while the authors were visiting the Institute for Pure and Applied Mathematics (IPAM), which is supported by the National Science Foundation. Susana Serna was supported by Spanish MINECO grant MTM2014-56218-C2-2-P. This work was performed under the auspices of the U.S. Department of Energy at the Lawrence Livermore National Laboratory under Contract No. DE-AC52-07NA27344.

\section{References}
\bibliography{scullard.bib}

\section{Appendix A: Derivation of the dielectric function}
Here, we compute the response function by computing the integral (\ref{eq:chi}) using our series expansion (\ref{eq:expansion}). First, we note that the response function is of the form
\begin{equation}
 \chi(k,\omega)=Z(k,\omega_-)-Z(k,\omega_+) \label{eq:chidiff}
\end{equation}
where
\begin{equation}
 Z(k,\omega) \equiv \lim_{\eta \rightarrow 0^+} \int d^3 {\bfv} \frac{f(\bfv)}{\hbar \omega - \hbar \bfv \cdot \bfk +i \eta} \label{eq:Z}
\end{equation}
and
\begin{equation}
\omega_\pm \equiv \omega \pm \frac{\hbar k^2}{2 m_e} \ .
\end{equation}
Equation (\ref{eq:chidiff}) can be easily found by making the substitution ${\bf u}={\bf v}+\hbar \bfk/m_e$ in the second term of (\ref{eq:chi}). Inserting the expansion (\ref{eq:expansion}) into (\ref{eq:Z}), we choose $\bfk$ to point in the $z$-direction and integrate in the cylindrical coordinates $(v_\perp,\phi,v_z)$. We then have
\begin{eqnarray}
& &Z(k,\omega)=\frac{n_e}{\hbar} \left(\frac{\beta m_e}{2 \pi}\right)^{\frac{1}{2}} \frac{1}{k} \sum A_k \times \cr 
& & \int_{-\infty}^\infty \int_0^\infty \frac{e^{-x-\beta m_e v_z^2/2} \Lkhalf \left(x+\frac{\beta m_e v_z^2}{2} \right)}{\omega/k-v_z + i \eta} dx d v_z
\end{eqnarray}
where we have made the substitution $x=m \beta v_\perp^2/2$. We now use the identity
\begin{equation}
\int_0^{\infty} e^{-x} \Lkhalf \left(x+ y \right) dx = \Lminhalf_k \left(y \right) \, \label{eq:sumint}
\end{equation}
which can be derived from the Laguerre sum formula
\begin{equation}
 L_n^{(\alpha_1+\alpha_2+1)}(x+y)=\sum_{i=0}^n L_i^{(\alpha_1)}(x) L_{n-i}^{(\alpha_2)}(y) , \label{eq:sumident}
\end{equation}
by choosing $\alpha_1=0$ and $\alpha_2=-1/2$. The integral we are left with is
\begin{equation}
Z(k,\omega)=\frac{n_e}{\hbar} \left(\frac{\beta m_e}{2 \pi}\right)^{\frac{1}{2}} \frac{1}{k} \sum A_m J_m
\end{equation}
where
\begin{equation}
J_m \equiv \int_{-\infty}^\infty \frac{e^{-\beta m_e v_z^2/2} \Lminhalf_m \left(\frac{\beta m_e v_z^2}{2} \right)}{\omega/k-v_z + i \eta} d v_z
\end{equation}
which, with the help of the substitution $x^2=\beta m_e v_z^2/2$, can be written
\begin{equation}
 J_m=\int_{-\infty}^\infty \frac{e^{-x^2}\Lminhalf_m(x^2)}{Y/\sqrt{2}-x+i \eta}dx \ .
\end{equation}
As usual, the imaginary part is easily found from the Sokhotski-Plemelj theorem,
\begin{equation}
 \lim_{\epsilon \rightarrow 0^+} \int_{-\infty}^\infty \frac{f(x)}{x \pm i \epsilon}=\mp i \pi f(0) + P \int_{-\infty}^\infty \frac{f(x)}{x}dx
\end{equation}
where $P$ denotes principal value integration. Thus,
\begin{equation}
 \mathrm{Im} J_m= - \pi e^{-Y^2/2} \Lminhalf_m \left(\frac{Y^2}{2} \right) \ .
\end{equation}
To find the real part, we will not directly attempt the principal value integral but will instead use the standard trick
\begin{equation}
 \frac{1}{\frac{Y}{\sqrt{2}}-x+i \eta}=-i \int_0^\infty e^{i(Y/\sqrt{2}-x+i\eta)t}dt
\end{equation}
and the identity
\begin{equation}
 \Lminhalf_m(x^2)=\frac{(-1)^m}{m! 2^{2m}}H_{2m}(x) \label{eq:Hermite}
\end{equation}
where $H_n(x)$ are Hermite polynomials, to write
\begin{eqnarray}
& &J_m=-i \frac{(-1)^m}{m! 2^{2m}} \times \cr 
& &\ \ \ \ \int_0^\infty \int_{-\infty}^\infty e^{-x^2} e^{i \left(\frac{Y}{\sqrt{2}}-x +i \eta \right)t } H_{2m}(x) dx dt \ .
\end{eqnarray}
Because $H_{2m}(x)$ and $e^{-x^2}$ are even, we have
\begin{eqnarray}
& &J_m=-i \frac{(-1)^m}{m! 2^{2m}} \times \cr 
& &\ \ \ \ 2 \int_0^\infty \int_{0}^\infty e^{-x^2} e^{i \left(\frac{Y}{\sqrt{2}} \right)t } \cos(xt) H_{2m}(x) dx dt \ .
\end{eqnarray}
Using 7.388.3 of Gradshteyn and Ryzhik \cite{Gradshteyn}, we find for the $x$-integral,
\begin{eqnarray}
& &\int_0^\infty e^{-x^2} \cos(xt) H_{2m}(x)dx= \cr
& &\ \ \ \ \ \ \ \ (-1)^m \frac{\sqrt{\pi}}{2} t^{2m} \exp \left(-\frac{t^2}{4}\right)
\end{eqnarray}
so
\begin{equation}
 \mathrm{Re} J_m=\frac{\sqrt{\pi}}{m! 2^{2m}} \int_0^\infty \sin \left(\frac{Y}{\sqrt{2}} t \right)  t^{2m} e^{-t^2/4} dt \ .
\end{equation}
Again consulting Gradshteyn and Ryzhik, this time 3.952.7, and using Kummer's transformation for the confluent hypergeometric function, $M(a,b;z)$, we arrive at the real part of $Z(k,\omega)$,
\begin{equation}
 \mathrm{Re}Z=n_e \frac{\beta m_e}{\hbar} \frac{\omega}{k^2} \sum_m A_m M \left(1+m,\frac{3}{2};-\frac{Y^2}{2} \right) \ .
\end{equation}
Putting together our previous results, the imaginary part is
\begin{equation}
 \mathrm{Im}Z=-\sqrt{\frac{\pi}{2}}n_e \frac{\sqrt{\beta m_e}}{\hbar k}\sum_m A_m e^{-\frac{Y^2}{2}} \Lminhalf_m \left(\frac{Y^2}{2} \right)
\end{equation}
Using (\ref{eq:chidiff}) and the definitions $Y_\pm^2 \equiv \beta m_e \omega_\pm^2/2 k^2$ and (\ref{eq:X}), we arrive at (\ref{eq:Qdelectric}). The calculation for arbitrary $u$ is the same except that one expands $H_{2m}(\sqrt{u}x)$ by means of a multiplication theorem for Hermite polynomials.

\section{Appendix B: Reduction of the collision integrals}
The purpose of this appendix is to derive a simplification of the quantum Lenard-Balescu collision integral using the polynomial expansion in equation (\ref{eq:expansion}). The calculation is only for $u=1$, which we use exclusively in this paper. The generalization to arbitrary $u$ is straightforward but results in more complicated formulas.

We multiply the left- and right-hand sides of the kinetic equation (\ref{eq:QLB}) by $\Lmhalf(\beta m_e v^2/2)$ and integrate over $d^3 {\bf v}$. The left-hand side becomes
\begin{equation}
 \frac{2 n_e}{\sqrt{\pi}}\frac{\Gamma(n+3/2)}{n!}\frac{d A_n}{dt} \ .
\end{equation}
The right-hand side is, of course, the real problem; it is the nine-fold integral
\begin{equation}
 \int C_{QLB}(f) \Lnhalf \left(\betavsq \right) d^3 {\bf v}
\end{equation}
which we will reduce to two. Beginning with ${\bf v'}$, we make use of the convenient definitions
\begin{eqnarray}
 \omega &\equiv& {\bf k} \cdot {\bf v}+\frac{\hbar k^2}{2 m_e} \\
 \omega_\pm &\equiv& \omega \pm \frac{\hbar k^2}{2 m_e} \
\end{eqnarray}
to write
\begin{eqnarray}
& &\int d^3 {\bf v'} \delta(\omega_+ - \bfk \cdot \bfvpr) f({\bf v'}) \cr
&=& n_e \left(\frac{\beta m_e}{2 \pi}\right)^{\frac{1}{2}} \frac{1}{k} \sum_{k=0}^{\infty} A_k(t) \Lminhalf_k \left(Y_+^2 \right) e^{-Y_+^2} \label{eq:vprime1}
\end{eqnarray}
and
\begin{eqnarray}
& &\int d^3 {\bf v'}  \delta(\omega_+ - \bfk \cdot \bfvpr) f(\bfvpr-\hbar \bfk/m_e) \cr
&=& n_e \left(\frac{\beta m_e}{2 \pi}\right)^{\frac{1}{2}} \frac{1}{k} \sum_{k=0}^{\infty} A_k(t) \Lminhalf_k \left(Y_-^2 \right) e^{-Y_-^2}, \label{eq:vprime2}
\end{eqnarray}
where
\begin{equation}
 Y_\pm^2 \equiv \frac{\beta m_e \omega_\pm^2}{2 k^2} \ .
\end{equation}
To derive (\ref{eq:vprime1}), we take ${\bf k}$ as the $z$-direction and integrate in the cylindrical coordinates $(v_\perp, \phi, v_z')$. We find then
\begin{eqnarray}
& &\int_{-\infty}^\infty \int_0^{2 \pi} \int_0^\infty \delta(\omega_+-k v_z') e^{-\frac{\beta m_e ({v'_\perp}^2+v_z'^2)}{2}} \cr
&\times& \Lnhalf \left(\frac{\beta m_e v_\perp^2}{2}+\frac{\beta m_e v_z'^2}{2} \right) v_\perp dv_\perp d \phi d v_z' \cr
&=& \frac{2 \pi}{\beta m_e} \frac{1}{k} e^{-Y_+^2} \int_0^{\infty} e^{-x} \Lnhalf \left(x+ Y_+^2 \right) dx \label{eq:cylint}
\end{eqnarray}
where we have made the subtitution $x=\beta m_e v_\perp^2 /2$. Combining (\ref{eq:sumint}) and (\ref{eq:cylint}) with the prefactors and series in $f(\bfvpr)$, we find (\ref{eq:vprime1}). Equation (\ref{eq:vprime2}) is found in exactly the same way after making the substitution ${\bf u}={\bf v}-\hbar {\bf k}/m_e$.

We aim in the end to have an integration over $k$ and $\omega$. For the ${\bf v}$ integration, we again take ${\bf k}$ to point in the $z$-direction and then we have $v_z=\omega_-/k$ and $dv_z=d \omega/k$. We will therefore employ cylindrical coordinates $(v_\perp,\phi,v_z)$ and integrate over $v_\perp$ and $\phi$, leaving $v_z$ as the $\omega$-integral. Because (\ref{eq:vprime1}) and (\ref{eq:vprime2}) depend only on $\omega$ and $k$ they will play no further role in the integration. The pieces we do need are
\begin{equation}
 \int d^3 {\bf v} f(\bfv) \Lnhalf \left(\betavsq \right)
\end{equation}
and
\begin{equation}
 \int d^3 {\bf v} f(\bfv+\hbar \bfk/m_e) \Lnhalf \left(\betavsq \right) \ ,
\end{equation}
integrated over $v_\perp$ and $\phi$, for which we find
\begin{eqnarray}
& &\int_0^{2 \pi} \int_0^\infty f({\bf v})  \Lnhalf \left(\betavsq \right) v_\perp d v_\perp d \phi = \cr
& &n_e \left(\frac{m_e \beta}{2 \pi}\right)^{3/2} 2 \pi \sum A_l(t) \int_0^\infty d v_\perp v_\perp e^{-\frac{\beta m_e (v_\perp^2+v_z^2)}{2}} \cr
&\times& \Lnhalf \left( \frac{\beta m_e v_\perp^2}{2}+Y_-^2 \right) L_l^{\left(\frac{1}{2}\right)} \left( \frac{\beta m_e v_\perp^2}{2}+Y_-^2 \right) \cr
&=& n_e \left(\frac{m_e \beta}{2 \pi}\right)^{1/2} e^{-Y_-^2} \sum A_l(t) \cr
&\times& \int_0^\infty e^{-x} \Lnhalf \left(x+Y_-^2 \right) \Llhalf \left(x+Y_-^2 \right) dx \label{eq:v1}
\end{eqnarray}
and, similarly
\begin{eqnarray}
& &\int_0^{2 \pi} \int_0^\infty f(\bfv+\hbar \bfk/m_e) \Lnhalf \left(\betavsq \right) v_\perp d v_\perp d \phi \cr
&=& n_e \left(\frac{m_e \beta}{2 \pi}\right)^{1/2} e^{-Y_+^2} \sum A_l(t) \cr
&\times& \int_0^\infty e^{-x} \Lnhalf \left(x+Y_-^2 \right) \Llhalf \left(x+Y_+^2 \right) dx \ . \label{eq:v2}
\end{eqnarray}
To handle the last integrals in (\ref{eq:v1}) and (\ref{eq:v2}) we use the identity
\begin{eqnarray}
& &\int_0^\infty e^{-x} \Lnhalf \left(x+y \right) \Llhalf \left(x+z \right) dx = \cr
& &\ \ \ \ \sum_{i=0}^{\min(l,n)} \Lminhalf_{n-i}(y) \Lminhalf_{l-i}(z)
\end{eqnarray}
which can easily be derived by again using (\ref{eq:sumident}) with $\alpha_1=0$ and $\alpha_2=-1/2$. At this point, only the magnitude of $k$ is left in the integrand, so $ \int d^3 {\bf k} \rightarrow 4 \pi \int k^2 dk$ . To get the final form of the integrand, we multiply (\ref{eq:vprime1}) by (\ref{eq:v1}),  subtract the product of (\ref{eq:vprime2}) and (\ref{eq:v2}), integrate over $4 \pi \int_{-\infty}^\infty dv_z \int_0^\infty k^2 dk$ and include the dimensional prefactor in (\ref{eq:Cee}). The resulting expression for the coefficients is
\begin{eqnarray}
& &C^n_{lk}=-\frac{\beta m_e n_e}{4 \pi^{3/2} \hbar^2}\frac{n!}{\Gamma(n+3/2)} \int_{-\infty}^\infty \int_0^\infty e^{-(Y_-^2+Y_+^2)} \cr
& &\ \ \ \ \ \ \ \ \ \times \frac{|\phi(k)|^2}{|\epsilon(k,\omega)|^2} q^n_{lk}(X,Y) dk d \omega
\end{eqnarray}
where $q^n_{lk}(X,Y)$ is given in equation (\ref{eq:q}). It is a straightforward matter to use the dimensionless variables $X$ and $Y$ defined in (\ref{eq:X}) and (\ref{eq:Y}) along with the Coulomb potential to arrive at (\ref{eq:Cnlk}).

\section{Appendix C: Exact integration over the dielectric function}
Here we will simplify equation (\ref{eq:Inlk}) by an exact integration over $X$. In fact, we will solve the more general case
\begin{equation}
 I_X(Y,s) \equiv \int_0^{\infty} dX  X^{2s-1} \frac{e^{-X^2}}{|\epsilon(X,Y)|^2}
\end{equation}
for which the $X$-integral in (\ref{eq:Inlk}) is the special case $s=0$. We do this because including terms in the collision integrals greater than $\mathrm{O}(1)$ requires integrals for which $s>0$, a generalization we may wish to consider in the future. These are also no more difficult than the $s=0$ case.

To begin, we make use of the identity
\begin{equation}
 \frac{1}{|\epsilon(X,Y)|^2}=\frac{1}{2 i \mathrm{Im}(\epsilon)} \left(\frac{1}{\epsilon*}-\frac{1}{\epsilon} \right) \ .
\end{equation}
Using the classical dielectric function in equation (\ref{eq:eps}), the integral becomes
\begin{eqnarray}
& &I_X(Y,s) =  \frac{1}{2 i \eta^2 w_i(Y)} \times \cr
& & \int_0^\infty \left[\frac{X^{2s+3} e^{-X^2}}{X^2+\eta^2 w^*(Y)}-\frac{X^{2s+3} e^{-X^2}}{X^2+\eta^2 w(Y)} \right] dX . \label{eq:IX}
\end{eqnarray}
Glancing at (\ref{eq:wr}) and (\ref{eq:wi}) it is clear that $w_i(Y)=-w_i(-Y)$ and $w_r(Y)=w_r(-Y)$ so that $w^*(Y)=w(-Y)$. The first term of (\ref{eq:IX}) can be written
\begin{equation}
  \frac{1}{2 i \eta^2 w_i(Y)} \int_0^\infty \frac{X^{2s+3} e^{-X^2}}{X^2+\eta^2 w(-Y)} dX .
\end{equation}
If in the $Y$-integration we make the substitution $Y \rightarrow -Y$, we find that the first and second terms of (\ref{eq:IX}) are actually equal and opposite when integrating over $Y$ and we can set
\begin{equation}
 I_X(Y,s) = -\frac{1}{i \eta^2 w_i(Y)} \int_0^\infty \frac{X^{2s+3} e^{-X^2}}{X^2+\eta^2 w(Y)} dX
\end{equation}
This integral can be evaluated in terms of special functions,
\begin{eqnarray}
& &\int_0^\infty \frac{X^{2s+3} e^{-X^2}}{X^2+\eta^2 w(Y)} dX = \frac{1}{2} e^{\eta^2 w(Y)} \times \cr
& &\left[\eta^2 w(Y) \right]^{1+s} (1+s)! \Gamma(-1-s,\eta^2 w(Y) ) \label{eq:exactX}
\end{eqnarray}
where $\Gamma(t,z)$ is the incomplete gamma function, defined by
\begin{equation}
 \Gamma(t,z) = \int_z^\infty u^{t-1} e^{-u} du \ .
\end{equation}
This function has the series,
\begin{eqnarray}
& &\Gamma(-1-s,z)=\frac{(-1)^{1+s}}{(1+s)!}[\psi_{s+2}-\gamma_E-\log z] \cr
&-&\frac{1}{z^{s+1}} \sum_{k=0, k \ne s+1}^\infty \frac{(-z)^k}{(k-s-1)k!} \ . \label{eq:gammaseries}
\end{eqnarray}
where $\gamma_E \approx 0.57721566$ is Euler's constant and $\psi_{s+2}$ are constants appearing in the digamma function at integer arguments. The first few of these are
\begin{eqnarray}
 \psi_2 &=& 1 \cr
 \psi_3 &=& 3/2 \cr
 \psi_4 &=& 11/6 \cr
 \psi_5 &=& 25/12 \ .\nonumber
\end{eqnarray}
The order-unity terms arising from dynamical screening come only from $s=0$ and we can now isolate these using (\ref{eq:gammaseries}) in (\ref{eq:exactX}) and expanding $\exp [\eta^2 w(Y)]$. Doing this, we find (\ref{eq:Inlk}).

\section{Appendix D: Exact expressions for Landau coefficients}
Here we derive equation (\ref{eq:exactSnlk}), the closed expression for $S^n_{lk}$, the coefficients of the Landau equation. We begin by setting $\epsilon(k,\omega)=1$ and using the variables $g \equiv k\sqrt{m_e \beta/2}$ and $z \equiv v_z\sqrt{m_e \beta/2}$, to write the coefficients for the quantum Landau equation $C^n_{lk}$,
\begin{eqnarray}
C^n_{lk} = - \bar{C} \sum_{j=0}^{\min(\ell,n)} \int_{g_0}^\infty dg \int_{-\infty}^{\infty} dz I_{QL}(g,z;\hbar)
\end{eqnarray}
where the integrand is
\begin{eqnarray}
& &I_{QL}(g,z;\hbar) \equiv \frac{1}{\hbar^2} \frac{1}{g^3} \exp\left( -z^2 - (z + \hbar g/m_e)^2 \right) \cr
& & \Lminhalf_{n-j} \left( z^2 \right) \left[ \Lminhalf_k \left( (z + \hbar g/m_e)^2 \right) \Lminhalf_{l-j} \left( z^2\right) \right. \cr
& & \ \ \ \ \ \ \ \ \ \ \ \ \ \ \ \left. - \Lminhalf_{\ell-j} \left( (z + \hbar g/m_e)^2 \right) \Lminhalf_k \left( z^2\right)\right],
\end{eqnarray}
the prefactor is now
\begin{equation}
 \bar{C}=\frac{\sqrt{\pi} n_e (2 m_e \beta)^{3/2} e^4 \, n!}{\Gamma(n+3/2)}
\end{equation}
and $g_0$ is the small-$k$ cutoff. To obtain the classical version of this expression, we take the $\hbar \rightarrow 0$ limit,
\begin{equation}
 I_{L}(g,z)=\lim_{\hbar \rightarrow 0} I_{QL}(g,z;\hbar)
\end{equation}
To do this we must expand to second-order in $\hbar$, and using some simple identities of Laguerre polynomials, this yields 
\begin{eqnarray}
& &I_{L}(g,z) = \frac{2 e^{-2 z^2} \Lminhalf_{n-j}(z^2)}{m_e^2 g} \times \cr
& &\Bigg[ (\ell-j+1) \Lminhalf_k (z^2) \Lminhalf_{\ell-j+1}(z^2) \cr 
& & \ \ \ \ \ \ - (k+1) \Lminhalf_{\ell-j}(z^2) \Lminhalf_k(z^2) \Bigg]
\end{eqnarray}
The $z$-integration of this expression can be done by taking advantage of the relationship between Laguerre and Hermite polynomials in equation (\ref{eq:Hermite}), and Titchmarsh's identity,
\begin{eqnarray}
& &\int_{-\infty}^\infty d z \; e^{-2z^2} H_a(z) H_b(z) H_c(z) = \frac{2^{(a+b+c-1)/2}}{\pi} \times \cr 
& & \Gamma\left( \frac{a+b-c+1}{2} \right) \Gamma\left( \frac{a-b+c+1}{2} \right) \times \cr 
& &\ \ \ \ \ \ \ \ \ \ \ \ \ \ \ \ \ \ \ \ \ \ \ \ \ \ \ \ \ \   \Gamma\left( \frac{-a+b+c+1}{2} \right) \ ,
\end{eqnarray}
when $a+b+c$ is even and the integral is zero otherwise. The coefficients then become
\begin{eqnarray}
& &C^n_{lk}= - \bar{C} \frac{\sqrt{2} \log\Lambda}{m_e^2 \pi}  \sum_{j=0}^{\text{min}(\ell,n)} (-1)^{n+k+\ell} \times \cr
& & \frac{2^{-(n+k+\ell-2j+1)}}{(n-j)! k! (\ell-j)!} \Gamma\left( -n+k+\ell+\frac{3}{2} \right) \times \cr
& &\left[ \Gamma\left( n+k-\ell-\frac{1}{2} \right)\Gamma\left( n-k+\ell-2j+\frac{3}{2} \right) \right. \cr
&-&\left. \Gamma\left( n+k-\ell+\frac{3}{2} \right) \Gamma\left( n-k+\ell-2j-\frac{1}{2} \right)\right]
\end{eqnarray}
At this point, we confess that we simply evaluated the above sum over $j$ in Mathematica, which returns an analytical form involving the regularized hypergeometric function, $_3\widetilde{F}_2$. Comparing this expression with equation (\ref{eq:LandauCnlk}) to get the correct numerical constants, we find equation (\ref{eq:exactSnlk}) for the $S^n_{lk}$. We do not yet know how to derive this formula legitimately, but we performed many checks between the analytic expression and the numerically-determined $S^n_{lk}$ to ensure that (\ref{eq:exactSnlk}) is indeed correct.

\section{Appendix E: Recurrence formulas for coefficients}
In this appendix, we derive and solve the recurrence formula for the triple product integrals $V^j_{lk}$ defined in equation (\ref{eq:V}). This will be based on the formula for Laguerre polynomials,
\begin{eqnarray}
 \Lminhalf_n(x)&=&\left(2-\frac{3/2+x}{n}\right) \Lminhalf_{n-1}(x) \cr
 & & \ \ \ \ \ \ \ \ - \left(1 -\frac{3}{2n}\right) \Lminhalf_{n-2}(x) \ . \label{eq:Lrecur}
\end{eqnarray}
Note that in the definition of $V$, two of the three Laguerre polynomials have the same argument. The strategy is to exploit this fact to find the recurrence formula for the family of integrands
\begin{equation}
 v^j_{lk}(X,Y) \equiv g(X,Y) \Lminhalf_j \left(Y_-^2 \right) \Lminhalf_k \left(Y_+^2 \right) \Lminhalf_l \left(Y_-^2 \right)
\end{equation}
where $g(X,Y)$ is an arbitrary function of $X$ and $Y$. We begin by using the Laguerre formula to show
\begin{eqnarray}
 & &\Lminhalf_j (Y_-^2) = \left(2-\frac{3}{2j} \right) \Lminhalf_{j-1}(Y_-^2) \cr
 &-&\left(1-\frac{3}{2j}\right) \Lminhalf_{j-2}(Y_-^2)-\frac{1}{j} Y_-^2 \Lminhalf_{j-1}(Y_-^2) \ .
\end{eqnarray}
Therefore,
\begin{eqnarray}
 & &v^j_{lk}(X,Y)=\left(2-\frac{3}{2j} \right) v^{j-1}_{lk}(X,Y) \cr
  &-& \left(1-\frac{3}{2j}\right)v^{j-2}_{lk}(X,Y)-\frac{1}{j}Y_-^2 v^{j-1}_{lk}(X,Y) . \label{eq:vrec}
\end{eqnarray}
We have a factor of $Y_-^2$ that will cause problems when we integrate. We can rid ourselves of it by using the recurrence formula again in the rearranged form
\begin{eqnarray}
 & &Y_-^2 \Lminhalf_{l}(Y_-^2)=\left(2 l +\frac{1}{2} \right)\Lminhalf_{l}(Y_-^2) \cr
 & &-\left(l-\frac{1}{2} \right)\Lminhalf_{l-1}(Y_-^2)-(l+1) \Lminhalf_{l+1}(Y_-^2)
\end{eqnarray}
to get
\begin{eqnarray}
 & &Y_-^2 v^{j-1}_{lk}(X,Y)=\left(2l+\frac{1}{2} \right) v^{j-1}_{lk}(X,Y) \cr
 &-& \left(l-\frac{1}{2} \right) v^{j-1}_{l-1,k}(X,Y) - (l+1) v^{j-1}_{l+1,k}(X,Y) 
\end{eqnarray}
which we can plug into (\ref{eq:vrec}) to get a recurrence formula for the integrand free of any additional factors of $X$ or $Y$. Putting everything together, we find
\begin{eqnarray}
V^j_{lk}&=&\left[2-\frac{2(l+1)}{j} \right] V^{j-1}_{lk} - \left(1-\frac{3}{2j}\right) V^{j-2}_{lk} \cr
&+& \frac{l-\frac{1}{2}}{j} V^{j-1}_{l-1,k}+\frac{l+1}{j}V^{j-1}_{l+1,k} \ . \label{eq:Vrecur}
\end{eqnarray}
Note that $k$ does not participate in this recurrence formula. This system is comprised of $\nmax+1$ discrete boundary-value problems in $j$ and $l$ corresponding to each possible value of $k$. For example, we can use our numerical integration techniques to find $V^j_{lk}$ for $j=0$ and $j=\nmax$ for all values of $k$. In principle, the recurrence formula then generates the rest. Note that although this is a three-point recurrence, it can be used to compute $j=1$ as long as we adhere to the convention that when $j>0$, we set negative-index polynomials to zero when they arise. This convention allows computation of the Laguerre polynomials themselves for $j>0$ and therefore also works for our coefficients. Another possibility is that rather than specify the values on the boundary of the cube, we instead compute the coefficients only for $j=0$. We now need to compute these up to $2 \nmax$. However, in practice it appears that this procedure is unstable and initial errors become out of control after a few iterations. We will not investigate this issue further here, but we point out that (\ref{eq:Vrecur}) can be solved exactly. To do this, rewrite (\ref{eq:Vrecur}) in the more symmetric form
\begin{eqnarray}
& &j V^j_{l-1,k}-2j V^{j-1}_{l-1,k}+\left(j-\frac{3}{2}\right) V^{j-2}_{l-1,k}= \cr
& & \ \ \ l V^{j-1}_{lk}-2j V^{j-1}_{l-1,k}+\left(j-\frac{3}{2}\right) V^{j-1}_{l-2,k}
\end{eqnarray}
so that $l-1$ is on the left and $j-1$ on the right. Looking for a separable solution of the form
\begin{equation}
 V^j_{lk}=U_j W_l \ 
\end{equation}
we find the equations
\begin{eqnarray}
 j U_j-2j U_{j-1}+\left(j-\frac{3}{2}\right) U_{j-2}&=&\lambda U_{j-1} \cr
 l W_l-2l W_{l-1}+\left(l-\frac{3}{2}\right) W_{l-2}&=&\lambda W_{l-1}
\end{eqnarray}
where $\lambda$ is an arbitrary constant. We can immediately recognize these as the recurrence formulas for the associated Laguerre polynomials with $\alpha=-1/2$ and $x=-3/2-\lambda$. The most general solution to (\ref{eq:Vrecur}) is a superposition of various values of $\lambda$ each with a different amplitude, $\mu$,
\begin{equation}
 V^j_{lk}=\sum_i \mu_i^{(k)} \Lminhalf_j \left(-\frac{3}{2}-\lambda^{(k)}_i \right) \Lminhalf_l \left(-\frac{3}{2}-\lambda^{(k)}_i \right) . \label{eq:Vexact}
\end{equation}
We require enough constants $\mu$ and $\lambda$ that we can satisfy the boundary conditions. It is not clear how best to apportion and calculate these but we appear to have several options. This requires further investigation, but it seems clear that if such an approach can work it would greatly improve the algorithm presented in the paper.

\end{document}